\documentclass[12pt]{article}

\usepackage{graphicx}

\def\rf#1{(\ref{eq:#1})}
\def\lab#1{\label{eq:#1}}
\def\nonu{\nonumber}
\def\br{\begin{eqnarray}}
\def\er{\end{eqnarray}}
\def\be{\begin{equation}}
\def\ee{\end{equation}}
\def\({\left(}
\def\){\right)}
\def\pa{\partial}
\def\rlx{\relax\leavevmode}
\def\IR{\rlx\hbox{\rm I\kern-.18em R}}
\def\vp{\varphi}
\newcommand{\sbr}[2]{\left\lbrack\,{#1}\, ,\,{#2}\,\right\rbrack}

\newcommand\bra[1]{\langle \, {#1}\, \mid}
\newcommand\ket[1]{\mid \, {#1} \, \rangle}

\def\IZ{\rlx\hbox{\sf Z\kern-.4em Z}}
\def\IR{\rlx\hbox{\rm I\kern-.18em R}}
\def\IC{\rlx\hbox{\,$\inbar\kern-.3em{\rm C}$}}

\def\FAaIA#1#2#3{{\sl Functional Analysis and Its Application} {\bf #1} (#2)
#3}

\def\PHSD#1#2#3{{\sl Physica} {\bf D#1} (#2) #3}
\def\PJA#1#2#3{{\sl Proc. Japan. Acad} {\bf #1A} (#2) #3}

\begin{document}

\begin{titlepage}
\vspace*{-1cm}

\vskip 3cm

\vspace{.2in}
\begin{center}
{\large\bf A simple formula for the conserved charges of soliton theories}
\end{center}

\vspace{.5cm}

\begin{center}
L. A. Ferreira~$^{1}$, and Wojtek J. Zakrzewski~$^{2}$

\vspace{.5 in}
\small

\par \vskip .2in \noindent
$^{(1)}$Instituto de F\'\i sica de S\~ao Carlos; IFSC/USP;\\
Universidade de S\~ao Paulo  \\ 
Caixa Postal 369, CEP 13560-970, S\~ao Carlos-SP, Brazil\\

\par \vskip .2in \noindent
$^{(2)}$~Department of Mathematical Sciences,\\
 University of Durham, Durham DH1 3LE, U.K.

\normalsize
\end{center}

\vspace{.5in}

\begin{abstract}

We present a simple formula for all the conserved charges of soliton
theories, evaluated on the solutions belonging to the orbit of the vacuum under
the group of 
dressing transformations. For pedagogical reasons we perform the
explicit calculations for the case of the sine-Gordon model, taken as
a prototype of soliton theories. We show that the energy and momentum are
boundary terms for all the solutions on the orbit of the vacuum. That
orbit includes practically all the solutions of physical interest,
namely solitons, multi-solitons, breathers, and combinations of
solitons and breathers. The example of the mKdV equation is also given
explicitly.

\end{abstract} 
\end{titlepage}

\section{Introduction}

\setcounter{equation}{0}

In the last years there appeared in the literature several results
pointing to the fact that the conserved charges, especially the
energy, of $1+1$ dimensional integrable field theories take the form
of boundary terms when evaluated on soliton solutions. In other words,
the charges seem to depend only on the asymptotic values of the fields
at infinity.  Using arguments based on conformal symmetry, it was
shown in \cite{acfgzhirota} that the energy of one-soliton solutions
of all abelian affine Toda theories is determined by the the asymptotic
value of a Belinfante type term (which `improves' the energy momentum
tensor). The 
same result was obtained in \cite{olivetensor} by explicit
calculations of the energy integral. By considering one of the light
cone variables as the time, it was shown in \cite{freeman} that the
corresponding chiral charges are surface terms. All these results
relied on the extensions of the Toda field theories proposed in
\cite{bb,afgz,cfgz}, by the addition of extra fields to render them
conformally invariant. Similar results were also obtained using
Backlund transformations \cite{ggsz}. 

In order to illustrate the statements above, consider the example of
the sine-Gordon model. Its conformal extension carried out along  the lines of
\cite{bb,afgz,cfgz} is defined by the eqs. of motion 
\br
\pa^2 \vp &=& - e^{\eta} \, \sin \vp, \nonu\\
\pa^2 \eta &=& 0, \nonu\\
\pa^2 \rho &=& e^{\eta}\,\(1- \cos \vp\).
\lab{sgeq}
\er

The theory is invariant under the conformal
transformations\footnote{The light cone coordinates $x_{\pm}$ are
  defined in \rf{lightcone}.} $x_{\pm}\rightarrow f_{\pm}\(x_{\pm}\)$ 
if the sine-Gordon field $\vp$ is a scalar under the conformal group
and if $e^{-\eta}\rightarrow f^{\prime}_{+}
f^{\prime}_{-}e^{-\eta}$. The conformal weights of $\rho$ are
arbitrary.  The Lagangrean for \rf{sgeq} is given by 
\be
{\cal L} = \frac{1}{2}\,
\(\pa_{\mu}\vp\)^2-\pa_{\mu}\eta\,\pa^{\mu}\rho -
e^{\eta}\,\(1-\cos\vp\) 
\ee
and the improved energy-momentum tensor by
\be
T_{\mu\nu}=\Theta_{\mu\nu}+ 2\(\pa_{\mu}\pa_{\nu}-g_{\mu\nu}\,\pa^2\)\, \rho,
\ee
where $\Theta_{\mu\nu}$ is the canonical energy-momentum tensor
\be
\Theta_{\mu\nu}= \pa_{\mu}\vp \, \pa_{\nu}\vp - \pa_{\mu}\rho \,
\pa_{\nu}\eta - \pa_{\mu}\eta \, \pa_{\nu}\rho - g_{\mu\nu}\, {\cal L}.
\lab{emtensor}
\ee
The second term in (3) is the above mentioned Belinfante type term
\cite{Belifante}. 

As a consequence of the conformal symmetry $T_{\mu\nu}$ is indeed
traceless.  The Hamiltonian of the pure sine-Gordon theory is
obtained by considering the solutions where the free field $\eta$ is a
constant (a spontaneous symmetry breaking of the conformal symmetry)
and is given by   
\be
{\cal H}_{SG} = \Theta_{00}\mid_{\eta = 0}= \frac{1}{2}
\(\pa_t\vp\)^2+ \frac{1}{2} \(\pa_x\vp\)^2+1-\cos\vp. 
\lab{sgham}
\ee
In \cite{acfgzhirota,olivetensor} it was shown that the
energy measured by the improved tensor, namely
$\int_{-\infty}^{\infty} dx\, T_{00}$, vanishes when evaluated on the
soliton solutions. 
Therefore, the energy measured by the sine-Gordon canonical energy
momentum tensor takes the form of a surface term, {\it i.e.}  
\be
E=\int dx\, {\cal H}_{SG} = - 2 \int_{-\infty}^{\infty}dx\, \pa_x^2 \,
\rho = -2 \pa_x \rho\mid_{x=-\infty}^{x=\infty}.
\lab{surfaceenergy} 
\ee

In this paper we extend this result by showing that not only the
soliton solutions but all solutions connected to the vacuum by the
so-called  dressing transformations, have the energy and momentum given
by the boundary terms \rf{surfaceenergy}. The orbit of solutions
obtained this way includes  practically all
solutions of physical interest like solitons, multi-solitons,
breathers, combinations of solitons and breathers, etc. We also give a
simple formula for the higher conserved charges for the same orbit of
solutions. For instance, for the case of $1$-soliton solutions we show that the
conserved charges take the form ($n=0,1,2,\ldots$)
\be
\Omega_{2n+1}^{(\pm)}= 
\pm 2\,\left[\frac{1+v}{1-v}\right]^{\pm\frac{\(2n+1\)}{2}},
\ee
where $v$ is the velocity of the soliton. For the breather solution
they are given by
\be
\Omega_{2n+1}^{(\pm)}= \pm
4\,\varepsilon\,\left[\frac{1+v}{1-v}\right]^{\pm\frac{\(2n+1\)}{2}}\,
\cos\left[\(2n+1\)\,\theta\right],
\ee
where again $v$ is the velocity of the breather, and the angle
$\theta$ is related to the breather oscillation frequency $\omega$ by $\sin
\theta = \omega$. For multi-soliton or multi-breather solutions, and
also for solutions that are combinations of solitons and breathers, all
the charges simply add up.

We also show that the reasons underlying such results
are not really connected to the conformal symmetry. They are a
consequence of very  
special algebraic structures appearing in the construction
of the solutions by the dressing transformation method. The first
important point is the behavior of the Wilson path ordered integral,
used in the construction of the conserved charges, under the dressing
transformation. One can write that integral in terms of its vacuum
value, which is simple, and the asymptotic values of the group element
performing the dressing (gauge) transformation. The second important
point relates to special decompositions of those group elements
involving oscillators algebras defined by the vacuum solution. We point out
however, that in order to prove our results one has to work with the
Kac-Moody algebra with a non-trivial central extension, even when the
soliton theory needs a zero curvature 
representation (Lax-Zakharov-Shabat equation)  
based on a loop algebra only. In many cases,
that implies the 
introduction of an extra field on the lines of \cite{bb,afgz,cfgz}.     

We also  extend  such results  to any integrable hierarchy
possessing the basic ingredients for the appearance of soliton
solutions. We work out the explicit formulas for all the conserved charges
for  such theories. The example of the modified
Korteweg-de Vries (mKdV) equation is given explicitly to ilustrate
that our method also works 
for non-Lorentz invariant theories appearing in fluid dynamics.  

The paper is organized as follows: in section
\ref{sec:conservedcharges} we discuss the construction of the
conserved charges of integrable theories using a flat connection
satisfying the Lax-Zakharov-Shabat equation, and show how the charges
relate to their vacuum value under the dressing transformations.  In
section \ref{sec:sg} we discuss in detail the case of the sine-Gordon
model as a prototype of soliton theories, and evaluate the  charges
explicitly. Section \ref{sec:general} is devoted to the generalization
of our results for any soliton theory satisfying the conditions given
at the beginning of that section. The example of the mKdV equation is
given in section \ref{sec:mkdv}. In appendix \ref{sec:sl2km} we give
some results about representation theory of Kac-Moody algebras needed in the
paper.

\section{The conserved charges}
\label{sec:conservedcharges}
\setcounter{equation}{0}

\begin{figure}
\scalebox{0.55}{
\includegraphics{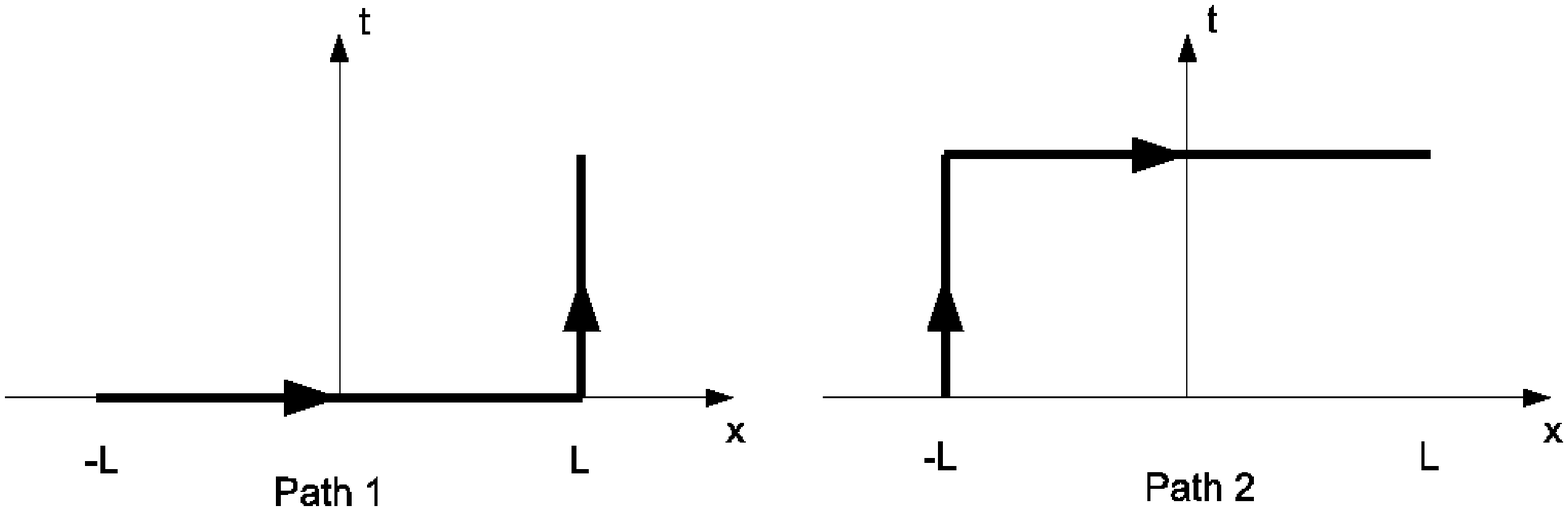}}
\caption{\label{fig:paths} }   
\end{figure}

A  $1+1$ dimensional integrable field theory admits a representation
of its equations of motion in terms of the so-called  zero curvature
condition, or Lax-Zakharov-Shabat equation \cite{zc} 
\be
F_{\mu\nu}=\pa_{\mu}A_{\nu} -\pa_{\nu}A_{\mu} + \sbr{A_{\mu}}{A_{\nu}}
= 0 \qquad \qquad \mu , \nu = 0,1,
\lab{zc}
\ee
where $A_{\mu}$ is a Lie algebra valued vector field which is a functional
of the physical fields of the theory. The vanishing of the curvature
$F_{\mu\nu}$ is equivalent to the equations of motion of the underlying field
theory. One of the key points of eq. \rf{zc} is that it constitutes
conservation laws in $1+1$ dimensions. The construction of the
corresponding conserved charges is obtained as follows:  Consider a
path ${\cal C}$ going from an initial point $P_0$ to a final point $P_1$, and
let the quantity $W$ be defined on ${\cal C}$ through the differential
equation 
\be
\frac{d\, W}{d\, \sigma} + A_{\mu}\, \frac{d\, x^{\mu}}{d\, \sigma}\, W=0,
\lab{wdef}
\ee
where $\sigma$ parametrizes ${\cal C}$. The solution of \rf{wdef} is the path
ordered integral  
\br
W&=& P\, e^{-\int_{C}d\sigma\, A_{\mu}\, \frac{d\, x^{\mu}}{d\,
    \sigma}}\nonu\\  
&=& \sum_{n=0}^{\infty} (-1)^n\,\int_0^{\sigma}d\sigma_1\,
\int_0^{\sigma_1}d\sigma_2\ldots \int_0^{\sigma_{n-1}}d\sigma_n \, 
A_{\mu_1}\(\sigma_1\) \frac{d\,x^{\mu_1}}{d\,\sigma_1}\ldots 
A_{\mu_n}\(\sigma_n\) \frac{d\,x^{\mu_n}}{d\,\sigma_n}.\nonu
\er
Eq. \rf{zc} is the sufficient condition for $W$  to be path
independent as long as the initial and end points of ${\cal C}$ are kept
fixed \cite{afs}. Then if we take the two paths shown in figure
\ref{fig:paths} we get that ($x^0\equiv t$, $x^1\equiv x$)  
\br
&& P\, \exp\(-\int_{0}^{t}dt \, A_t\mid_{x=L}\)\, P\,
\exp\(-\int_{-L}^{L}dx \, A_x\mid_{t=0}\) = \nonu\\ 
&=&P\, \exp\(-\int_{-L}^{L}dx \, A_x\mid_{t=t}\)\, P\,
\exp\(-\int_{0}^{t}dt \, A_t\mid_{x=-L}\).\nonu 
\er

Next we impose the boundary condition (for $L\rightarrow \infty$)
\be
A_t\mid_{x=L} = A_t\mid_{x=-L} + \beta \, C
\lab{bc}
\ee
where $\beta$ is some function of $t$ and $L$
and $C$ is the central charge of the algebra. Then 
one gets the quasi-isospectral evolution equation 
\be
W_t = e^{\int_0^t dt\, \beta \, C} \; U\(t\)\, W_0 \, U\(t\)^{-1},
\ee
where
\be
W_{0/t} = P\, \exp\(-\int_{-L}^{L}dx \, A_x\mid_{t=0/t}\)\; ;  \qquad 
U\(t\)= P\, \exp\(-\int_{0}^{t}dt \, A_t\mid_{x=L}\).
\lab{wudef}
\ee
If $\beta$ vanishes we have a pure isospectral evolution. Therefore,
the eigenvalues of $W_t$ are constant in time, and constitute the
conserved charges of the underlying field theory associated with \rf{zc}.  However,
if $\beta$ does not vanish we still can have conserved charges in some
circumstances. For instance, suppose that the operator $\Psi_0$ is an
eigenstate of $W_0$ under the adjoint action 
\be
W_0\, \Psi_0 \, W_0^{-1} = \lambda \, \Psi_0.
\lab{eigenvalueeq}
\ee
Then, the operator $\Psi_t = U\(t\)\, \Psi_0\, U\(t\)^{-1}$ is an
eigenstate of $W_t$ with the same eigenvalue, since the central term
$C$ commutes with every operator.  

Another key point of eq. \rf{zc} is that it is invariant under the
gauge transformations 
\be
A_{\mu} \rightarrow A_{\mu}^g= g\, A_{\mu}\, g^{-1} - \pa_{\mu}g\, g^{-1},
\lab{gauge}
\ee
where $g$ is an element of the Lie group associated to the Lie algebra
of $A_{\mu}$. Under \rf{gauge} the path ordered integral
transforms as 
\be
W\({\cal C}\) \rightarrow W^g\({\cal C}\)=g\(P_1\)\, W\({\cal C}\)\,
g^{-1}\(P_0\), 
\lab{wgauge}
\ee
where $P_0$ and $P_1$ are the initial and final points of ${\cal C}$
respectively. One then observes that the conserved charges are
invariant under those gauge transformations for which $g\(t,x=-L\) =
e^{\alpha\, C}\, g\(t, x=L\)$, since if $\Psi_t$ is an eigenvector of
$W_t$ under the adjoint action, {\it i.e.} $W_t\, \Psi_t\, W_t^{-1}=\lambda
\, \Psi_t$, so is $\Psi_t^g = g\(t, x=L\)\, \Psi_t\, g\(t,
x=L\)^{-1}$, an eigenvector of $W_t^g= e^{-\alpha\, C}\,g\(t,x=L\)\,
W_t\, g^{-1}\(t, x=L\)$. However, we are more interested in the
gauge transformations that do change the values of the conserved
charges, as we explain below.

The transformations \rf{gauge} constitute the so-called hidden
symmetries of the underlying field theory associated with \rf{zc}, in the sense
that they are not symmetries of the equations of motion but of the
zero curvature condition. Under some circumstances, the
transformations \rf{gauge} constitute a map among solutions of the
theory. In fact, all the soliton and multi-soliton solutions in $1+1$
dimensions can be constructed using special transformations of the type
\rf{gauge}, named dressing transformations, starting from a simple
vacuum solution.  
Therefore, if one knows the operator $W_t^{({\rm vac.})}$ associated
to a given simple vacuum solution, and knows the dressing
transformation that maps that vacuum solution to a non-trivial
solution, like a soliton, then the corresponding operator will be
given by 
\be
W_t = g\(t,x=L\)\, W_t^{({\rm vac.)}}\, g^{-1}\(t, x=-L\).
\lab{wtnew}
\ee
Consequently, the conserved charges evaluated on such non-trivial
solutions, which are the eigenvalues of $W_t$, will depend upon the
eigenvalues of $W_t^{({\rm vac.)}}$, which are trivial, and on the
asymptotic values of the group element performing the
transformation. In many cases, that will imply that the charges are
surface terms as we now explain on some concrete examples.

\section{The case of the sine-Gordon model}
\label{sec:sg}
\setcounter{equation}{0}

The standard zero curvature condition \rf{zc} for the sine-Gordon
model involves potentials $A_{\mu}$ which live in a $sl(2)$ loop
algebra, {\it i.e.}, they are $2\times 2$ matrices depending on a so-called
spectral parameter. That is an infinite dimensional Lie algebra
without a central extension. The potentials are given by 
\br
A_{+}&=& \frac{1}{2}\(
\begin{array}{cc}
0 & e^{i\vp}\\
\lambda\, e^{-i\vp}&0
\end{array}\)=
\frac{1}{2}\,\( \cos \vp \;  b_1 + i\, \sin \vp \; F_1\),
\nonu\\
A_{-}&=& -\frac{1}{2}\(
\begin{array}{cc}
i\,\partial_{-}\vp & 1/\lambda\\
1 & -i\,\partial_{-}\vp
\end{array}\)=
- \frac{1}{2}\, b_{-1} - \frac{i}{2}\,\pa_{-}\vp \, F_0, 
\lab{aa}
\er
where the $2\times 2$ matrix representation for $b_{\pm 1}$, $F_0$ and
$F_1$ are given in \rf{matrixrep} and $\lambda$ is the so-called spectral
parameter. Moreover, in \rf{aa}  we have used light cone coordinates 
\be
x_{\pm}= \frac{1}{2}\( t \pm x\) \quad \qquad \pa_{\pm} =  \pa_t\pm
\pa_x \quad \qquad \pa^2=\pa_t^2-\pa_x^2=\pa_{+}\pa_{-}. 
\lab{lightcone}
\ee
Putting \rf{aa} into the zero curvature condition \rf{zc} one finds
that the diagonal part of the matrices  gives the sine-Gordon equation
\be
\pa^2 \vp = -  \sin \vp 
\ee
and the off-diagonal part is satisfied trivially.

However, to present our arguments that demonstrate the existence of conserved
charges we need to centrally  extend the basic algebra. We
will then work with the full $sl(2)$ Kac-Moody algebra. In order for
the zero curvature to remain valid on such algebra, it is necessary to
extend the  sine-Gordon model by  the addition of an extra scalar field.
Furthermore,
for the theory to possess a Lagrangian, we need to add a further
scalar field, which in fact renders the model conformally
invariant. This way we end up with the so-called conformal sine-Gordon model
\cite{bb,afgz,cfgz} defined by the equations of motion \rf{sgeq}.  

The three equations \rf{sgeq} are equivalent to \rf{zc} with the
potentials $A_{\mu}$ given by  
\br
A_{+}&=& \frac{1}{2}\,e^{\eta}\,\( \cos \vp \;  b_1 + i\, \sin \vp \;
F_1\), \nonu\\ 
A_{-}&=& - \frac{1}{2}\, b_{-1} - \frac{i}{2}\,\pa_{-}\vp \, F_0 
- \pa_{-}\eta \, Q -
\frac{1}{4}\,\pa_{-}\(\rho + \gamma\)\, C,  
\lab{sgpot}
\er
where $\gamma$ is a function satisfying 
\be
\pa_{+}\pa_{-}\gamma= -e^{\eta}
\lab{gammadef}
\ee 
and where $F_0$, $F_1$, $b_{\pm 1}$ and $C$ are generators of a $sl(2)$
Kac-Moody algebra \cite{goreview}. Its generators and commutation
relations are 
defined on the appendix \ref{sec:sl2km}. This time, {\it i.e.} for
$C\ne0$,  
we do not have a finite matrix representation of the algebra and we
have to proceed 
using just the commutation relations given in the appendix \ref{sec:sl2km}. 

However, we are really interested only in finite energy solutions of
the pure sine-Gordon 
model ($\eta =0$). From \rf{sgham} one sees that the finiteness of the
energy requires  that  $\vp\(t,x=\pm L\) \rightarrow 2 \pi n_{\pm}$,
with $n_{\pm}$  integers,  for $L\rightarrow \infty$, but this does not
impose any condition on the behaviour of the $\rho$ field. Therefore,
the potentials \rf{sgpot} can satisfy the boundary condition \rf{bc} since 
\br
A_t\(t,x=L\) &=& A_t\(t,x=-L\) - \frac{1}{8}\,
\pa_{-}\(\rho+\gamma\)\mid_{x=-L}^{x=L}\, C \nonu\\ 
&=& \frac{1}{4}\,\(b_1-b_{-1}\) - 
\frac{1}{8}\, \pa_{-}\(\rho+\gamma\)\mid_{x=L}\, C.
\er
In consequence, the conserved charges can be constructed as explained in
\rf{bc}-\rf{eigenvalueeq}. In order to do this we have to build the operator
$W_t$ from its form for a vacuum configuration as explained in
\rf{wtnew}. Note that the conformal sine-Gordon eqs. \rf{sgeq} have
a vacuum  solution given by $\vp=\eta=\rho=0$. The potentials
$A_{\pm}$ given  in \rf{sgpot},  when evaluated on such a vacuum
solution become
\br
A^{({\rm vac.})}_{+} &=& \frac{1}{2}\, b_{1},  \nonu\\
A^{({\rm vac.})}_{-} &=& - \frac{1}{2}\, b_{-1} - \frac{1}{4}\,\pa_{-}
\gamma^{({\rm vac.})}\, C ,
\lab{sgpotvac}
\er
where, according to \rf{gammadef}, $\pa_{+}\pa_{-} \gamma^{({\rm
    vac.})}=-1$, and so  $\gamma^{({\rm vac.})}=-x_{+}\, x_{-}$. Since
these potentials are flat we can write them as 
\be
A^{({\rm vac.})}_{\mu}= - \partial_{\mu} \Psi_{\rm vac}\,\Psi_{\rm
  vac}^{-1}
\ee
with
\be
 \Psi_{\rm vac} = e^{-\frac{1}{2}\, x_{+}\, b_1}\, 
e^{\frac{1}{2}\, x_{-}\, b_{-1}}.
\lab{psivac}
\ee
The solutions we are interested are those in the orbit of such a vacuum
solution under the group of the so-called dressing transformations
\cite{dressing}. In order to construct such an orbit of solutions we
consider a constant group element $h$, obtained by exponentiating the
generators of the $sl(2)$ Kac-Moody algebra, which admit the following
Gauss like decomposition
\be
\Psi_{\rm vac}\,h\,\Psi_{\rm vac}^{-1} = G_{-}^{-1}\, G_0^{-1}\, G_{+},
\lab{gaussdecomp}
\ee
where $G_{+}$, $G_{-}$, and $ G_0$ are group elements obtained by
exponentiating the generators of the positive, negative and zero grades
respectively, of the grading operator $Q$ defined in \rf{graddef}. 

Then we define the potential
\be
A^h_{\mu}= -\partial_{\mu}\Psi_h\, \Psi_h^{-1}
\ee
with
\be
\Psi_h = G_0\,G_{-}\, \Psi_{\rm vac}\,h= G_{+}\, \Psi_{\rm vac}.
\lab{dresspsi}
\ee
As a consequence we have
\br
A^h_{\mu}&=& G_{+}\,A^{({\rm vac.})}_{\mu}\,G_{+}^{-1}- \partial_{\mu}
G_{+}\, G_{+}^{-1}
\lab{dress1}\\
&=& G_0\(G_{-}\,A^{({\rm vac.})}_{\mu}\,G_{-}^{-1}- \partial_{\mu}
G_{-}\, G_{-}^{-1}\)G_0^{-1}-\partial_{\mu}
G_{0}\, G_{0}^{-1}.
\lab{dress2}
\er

The fact that $A^h_{\mu}$ and $A^{({\rm vac.})}_{\mu}$ are related by
two gauge transformations, one in\-vol\-ving only positive grade
generators and the other only non-positive grade generators,
guarantees that $A^h_{\mu}$ has the same grading structure as $A^{({\rm
    vac.})}_{\mu}$, and so as $A_{\mu}$ defined in
\rf{sgpot}. Indeed, the $x_{+}$-component of \rf{dress1} implies that
$A^h_{+}$ has components of grades 
greater than or equal to one, and the $x_{+}$-component of \rf{dress2}
implies that it has 
components of grades smaller or equal to one. Thus, $A^h_{+}$
must have components of grade one only. The same reasoning applies to
$A^h_{-}$, using the $x_{-}$-components of \rf{dress1} and \rf{dress2}.
Notice that the 
space-time dependency of $A^h_{\mu}$  is explicit, since it
depends on the parameters of $G_{0,\pm}$ which, according to
\rf{gaussdecomp},  are explicit functions of the space-time
variables. Therefore $A^h_{\mu}$ corresponds to $A_{\mu}$ evaluated on
the solution constructed by the dressing method. By equating 
$A^h_{\mu}$ to  $A_{\mu}$ one then generates an explicit solution for the
fields, since $A_{\mu}$ is their functional. Note that the
dressing transformation from a vacuum with 
$\eta=0$ will never produce a solution with $\eta\neq 0$. The reason for this 
is that the grading operator $Q$ can never be obtained as a result of any
commutator and, consequently, the terms proportional to $Q$ will never
appear in \rf{dress1}-\rf{dress2}. 

The best way of extracting the solutions for the fields is as
follows:  first note that the
grade zero part of the $x_{-}$-component of \rf{dress2} is
$\(-\partial_{-}G_{0}\, 
G_{0}^{-1}- \frac{1}{4}\,\pa_{-} \gamma^{({\rm vac.})}\, C\)$. Comparing
this with the grade zero component of $A_{-}$ in \rf{sgpot} (with
$\eta=0$ since 
the dressing transformation does not excite $\eta$) one gets that  
\be
G_0=e^{\frac{i}{2}\, \vp\, F_0+\frac{1}{4}\rho \, C}.
\ee

The highest weight states of the two fundamental representations of
the $sl(2)$ Kac-Moody algebra are annihilated by the positive grade
generators and so one has \hfil
$G_{+}\, \ket{\lambda_i}= \ket{\lambda_i}$,
and $\bra{\lambda_i} G_{-}=  \bra{\lambda_i}$, for $i=0,1$. Then,
using the relations \rf{highestweight1}-\rf{highestweight2} one gets
from \rf{gaussdecomp} that 
\br
\tau_0 &\equiv& \bra{\lambda_0} \Psi_{\rm vac}\,h\,\Psi_{\rm
  vac}^{-1}\ket{\lambda_0} = \bra{\lambda_0} G_0^{-1}\ket{\lambda_0} = 
e^{\frac{i}{4}\, \vp -\frac{1}{4}\rho },\nonu\\
\tau_1 &\equiv& \bra{\lambda_1} \Psi_{\rm vac}\,h\,\Psi_{\rm
  vac}^{-1}\ket{\lambda_1} = \bra{\lambda_1} G_0^{-1}\ket{\lambda_1} = 
e^{-\frac{i}{4}\, \vp -\frac{1}{4}\rho }
\lab{taudef1}
\er
and so
\be
\vp =- 2\, i\, \log \frac{\tau_0}{\tau_1},\qquad \quad 
\rho =- 2 \, \log \( \tau_0\,\tau_1\).
\lab{taudef2}
\ee

In fact, the highest weight states are eigenvectors of $G_0^{-1}$
\be
G_0^{-1}\, \ket{\lambda_i}= \tau_i\,\ket{\lambda_i}, \qquad \qquad 
\bra{\lambda_i}\,G_0^{-1}=  \tau_i\,\bra{\lambda_i}, \quad\quad \qquad
i=0,1.
\lab{g0eigenvalue}
\ee
The quantities $\tau_0$ and $\tau_1$ are the so-called Hirota's tau
functions. Indeed, substituting \rf{taudef2} into \rf{sgeq} one finds
that they satisfy the Hirota's equations
\br
\tau_0\,\partial_{+}\partial_{-}\tau_0-\partial_{+}\tau_0\,\partial_{-}\tau_0
&=& \frac{1}{4}\(\tau_0^2-\tau_1^2\),\nonu\\
\tau_1\,\partial_{+}\partial_{-}\tau_1-\partial_{+}\tau_1\,\partial_{-}\tau_1
&=& \frac{1}{4}\(\tau_1^2-\tau_0^2\).
\lab{taueq}
\er

We now write the group elements $G_{\pm}$ as
\be
G_{\pm}= g_{\pm ,F}^{-1}\, g_{\pm ,b}, \qquad  
g_{\pm ,F}=\exp\(\sum_{n=1}^{\infty} \zeta_n^{(\pm )} \, F_{\pm n}\), \quad
g_{\pm ,b}=\exp\(\sum_{n=0}^{\infty} \xi_{2n+1}^{(\pm)} \, b_{\pm
  (2n+1)}\).
\lab{gpmdef}
\ee
The relation \rf{dress1}  can then be rewritten as
\be
g_{+,F}\,A_{\mu}^h\,g_{+,F}^{-1}-\partial_{\mu}g_{+,F}\,g_{+,F}^{-1} = 
g_{+,b}\,A^{({\rm
    vac.})}_{\mu}\,g_{+,b}^{-1}-\partial_{\mu}g_{+,b}\,g_{+,b}^{-1}\equiv 
a_{\mu}^{(+)}  
\lab{nicerel1}
\ee
and the relation \rf{dress2}  as
\be
g_{-,F}\,{\bar
  A}_{\mu}^h\,g_{-,F}^{-1}-\partial_{\mu}g_{-,F}\,g_{-,F}^{-1} =  
g_{-,b}\,A^{({\rm
    vac.})}_{\mu}\,g_{-,b}^{-1}-\partial_{\mu}g_{-,b}\,g_{-,b}^{-1}\equiv 
a_{\mu}^{(-)},  
\lab{nicerel2}
\ee
which serve as the definitions of the potentials $a_{\mu}^{(+)}$ and
$a_{\mu}^{(-)}$,  
and the potential ${\bar A}_{\mu}^h$ is defined as
\be
{\bar A}_{\mu}^h\equiv G_0^{-1}\,A_{\mu}^h\,G_0-\partial_{\mu}G_0^{-1}\,G_0 .
\ee
The gauge transformations relating the various potentials can  be
summarized in the following diagrams 
\br
 \begin{array}{ccc}
A^{({\rm vac.})}_{\mu} & \stackrel{G_{+}}{\longrightarrow} & A_{\mu}^h\\
 & \vcenter{\llap{\footnotesize{$g_{+,b}$}}}\searrow &
 \Big\downarrow\vcenter{\rlap{\footnotesize{$g_{+,F}$}}}\\
& & a_{\mu}^{(+)}
\end{array} \qquad\qquad\qquad\qquad
\begin{array}{ccc}
A^{({\rm vac.})}_{\mu} & \stackrel{G_{-}}{\longrightarrow} & {\bar
 A}_{\mu}^h\\ 
 & \vcenter{\llap{\footnotesize{$g_{-,b}$}}}\searrow &
 \Big\downarrow\vcenter{\rlap{\footnotesize{$g_{-,F}$}}}\\
& & a_{\mu}^{(-)}
\end{array}.
\lab{diagram}
\er
The potentials $a_{\mu}^{(+)}$ and $a_{\mu}^{(-)}$ are related to the
abelian potentials considered in \cite{afgzcharges,oliveabelian}

Equating $A_{\mu}^h$ to $A_{\mu}$, given in \rf{sgpot}, with $\eta
=0$, we get that 
\br
{\bar A}_{+}^h &=& \frac{1}{2} \, b_1 + \frac{i}{2}\,
\partial_{+}\vp\, F_0 + \frac{1}{4}\, \partial_{+}\rho\, C,\nonu\\
{\bar A}_{-}^h &=& -\frac{1}{2}\, \( \cos \vp \; b_{-1}-i\, \sin\vp\;
F_{-1} \) - \frac{1}{4}\partial_{-} \gamma^{({\rm vac.})}\, C.
\er
In addition, the $x_{-}$ component of \rf{nicerel1} gives
\br
g_{+,F}\,\(- \frac{1}{2}\, b_{-1} - \frac{i}{2}\,\pa_{-}\vp \,
F_0\)\,g_{+,F}^{-1} -\frac{1}{4}\,\pa_{-}\(\rho + \gamma^{({\rm
    vac.})}\)\, C   
-\partial_{-}g_{+,F}\,g_{+,F}^{-1}  \nonu\\
= \, - \frac{1}{2}\, b_{-1} 
-\frac{1}{4}\,\( 2\, \xi_1^{(+)}+\pa_{-}\gamma^{({\rm vac.})}\)\, C
-\sum_{n=0}^{\infty} \partial_{-}\xi_{2n+1}^{(+)} \, b_{2n+1}
\lab{goodcond1}
\er
and the $x_{+}$ component of \rf{nicerel2} gives
\br
g_{-,F}\,\( \frac{1}{2}\, b_{1} + \frac{i}{2}\,\pa_{+}\vp \,
F_0\)\,g_{-,F}^{-1} +\frac{1}{4}\,\pa_{+}\rho \, C  
-\partial_{+}g_{-,F}\,g_{-,F}^{-1}  \nonu\\
= \,  \frac{1}{2}\, b_{1} 
-\frac{1}{2}\, \xi_1^{(-)}\, C
-\sum_{n=0}^{\infty} \partial_{+}\xi_{2n+1}^{(-)} \, b_{-2n-1}.
\lab{goodcond2}
\er

Observe that the r.h.s. of \rf{goodcond1} and \rf{goodcond2} contain
terms proportional to the oscillators $b_{2n+1}$ and to the central
term $C$ only. Therefore, the components on the l.h.s. of these
equations, which are in the direction of the $F_n$'s must vanish. Splitting the
relations \rf{goodcond1} and \rf{goodcond2} into eigenvectors of the
grading operator $Q$ one can then determine the parameters of $g_{\pm
  ,F}$ recursively. Indeed, one finds that
\br
\zeta_1^{(+)}&=& -\frac{i}{2}\, \partial_{-}\vp ,\qquad\qquad \quad 
\zeta_1^{(-)}= -\frac{i}{2}\, \partial_{+}\vp ,\nonu\\
\zeta_2^{(+)}&=& \frac{i}{2}\, \partial_{-}^2\vp, \qquad\qquad \qquad 
\zeta_2^{(-)}= -\frac{i}{2}\, \partial_{+}^2\vp,\\
&\vdots& \qquad\qquad \qquad\qquad\qquad \quad \vdots\nonu
\er
So, $\zeta_n^{(\pm)}$ are polynomials in the $x_{\mp}$ derivatives of
the field $\vp$ and they do not depend on the field $\rho$. As
discussed above, for finite energy solutions, one needs $\vp
\rightarrow 2 \pi n_{\pm}$ as $x\rightarrow \pm \infty$, 
with $n_{\pm}$  integers, and consequently
\be
g_{\pm ,F} \rightarrow 1 \qquad {\rm for } \quad  x\rightarrow \pm
\infty.
\lab{bcgpmf}
\ee
We also get from \rf{goodcond1} and \rf{goodcond2} that
\br
\xi_1^{(+)}&=& \frac{1}{2}\, \partial_{-}\rho, \qquad \qquad \qquad \quad \;\;
\xi_1^{(-)}= - \frac{1}{2}\, \partial_{+}\rho, \nonu\\
\partial_{-}\xi_1^{(+)}&=& - \frac{1}{4}\, \(\partial_{-}\vp\)^2,
\qquad \qquad 
\partial_{+}\xi_1^{(-)}=  \frac{1}{4}\, \(\partial_{+}\vp\)^2,
\lab{importantrel1}\\
&\vdots& \qquad \qquad \qquad\qquad \qquad \qquad \;\; \vdots \nonu
\er

From the relations \rf{importantrel1} we obtain an important property
of the solutions in the orbit of the vacuum,  which may not
necessarily hold for other solutions of \rf{sgeq}. To get it we note that
\rf{importantrel1} implies that 
\be
\partial_{-}^2\rho = - \frac{1}{2}\, \(\partial_{-}\vp\)^2, \qquad
\qquad \qquad
\partial_{+}^2\rho = - \frac{1}{2}\, \(\partial_{+}\vp\)^2 .
\lab{importantrel2}
\ee
Using \rf{importantrel2}, \rf{lightcone}, and the third eq. of
\rf{sgeq} we see that the components of the canonical energy momentum tensor
\rf{emtensor}, for $\eta=0$ (see \rf{sgham}), can be written as  
\br
\Theta_{00}\mid_{\eta = 0}&=& \frac{1}{2} \(\pa_t\vp\)^2+ \frac{1}{2}
\(\pa_x\vp\)^2+1-\cos\vp = -2\, \partial_x^2\rho, \nonu\\
\Theta_{01}\mid_{\eta = 0}&=& \partial_t\vp\; \partial_x\vp = -2\, 
\partial_t\partial_x\rho. 
\er
In consequence, the energy and momentum of the solutions on the orbit of
the vacuum are surface terms:
\br
E&\equiv& \int_{-\infty}^{\infty}dx\, \Theta_{00}\mid_{\eta = 0} = 
-2\, \partial_x\rho\mid_{x=-\infty}^{x=\infty},\nonu\\
P&\equiv& \int_{-\infty}^{\infty}dx\, \Theta_{01}\mid_{\eta = 0} =
-2\, \partial_t\rho\mid_{x=-\infty}^{x=\infty}.
\lab{energy-momentum-formula}
\er
Replacing \rf{taudef2} into \rf{importantrel2} one gets that the
$\tau$-functions, evaluated on the solutions on the orbit of the
vacuum, satisfy, in addition to the Hirota's eqs. \rf{taueq}, the relations 
\be
\tau_1\partial^2_{\pm}\tau_0 + \tau_0\partial^2_{\pm}\tau_1 - 2\,
\partial_{\pm}\tau_0\, \partial_{\pm}\tau_1 = 0
\label{nicerels}
\ee

The infinite number of conserved charges for the sine-Gordon model can
be easily derived using the arguments of section
\ref{sec:conservedcharges} and the gauge transformations defined in
this section. 
Thus, from the definition \rf{wudef} we see that the $W_t$ operator for the
vacuum potential \rf{sgpotvac} is given by 
\be
W_t^{({\rm vac.})} = e^{-\frac{L}{2}\, b_{-1}}\; e^{-\frac{L}{2}\,
  b_{1}}\; e^{\frac{1}{8}\( L^2 + \int_{-L}^L dx\,
  \pa_{-}\gamma^{({\rm vac.})}\)\; C} . 
\ee
Next we define the operators:
\be
\Psi_{2n+1} = \; : e^{ b_{2n+1}+ b_{-2n-1}}:\;\;  =  
e^{ b_{-2n-1}}\; e^{ b_{2n+1}} 
\qquad \qquad \qquad n=0,1,2,\ldots
\lab{psidef}
\ee
where $::$ stands for the normal ordering of the oscillators $b_{2n+1}$. We
see that, under the adjoint action, $\Psi_{2n+1}$ are eigenvectors of
$W_t^{({\rm vac.})}$,  with unity eigenvalue  since 
\be
W_t^{({\rm vac.})} \, \Psi_{2n+1}\, {W_t^{({\rm vac.})}}^{-1} =  \Psi_{2n+1}.
\ee
Therefore, the conserved charges for the vacuum solution are indeed
trivial. 

Notice from \rf{diagram} and \rf{bcgpmf} that the gauge potentials
$A_{\mu}^h$ and $a_{\mu}^{(+)}$, and also ${\bar A}_{\mu}^h$ and
$a_{\mu}^{(-)}$, are connected by gauge transformations involving group
elements, namely $g_{\pm,F}$, that go to unity at spatial
infinity. Therefore, from the arguments given below \rf{wgauge}, one
concludes that the conserved charges constructed from $A_{\mu}^h$ and
$a_{\mu}^{(+)}$ are the same. For the same reasons, the charges obtained
from ${\bar A}_{\mu}^h$ and $a_{\mu}^{(-)}$, are also equal. We can
then construct the charges  from the potentials $a_{\mu}^{(+)}$ and
$a_{\mu}^{(-)}$, because the calculations are easier since these potentials are
related to the vacuum potential via abelian gauge transformations. 

The $W_t$ operators for the non-trivial solutions connected to the
vacuum by the gauge transformations performed by the group elements
$g_{\pm,b}$, according to \rf{wtnew}, are given by 
\be
W_t^{(\pm)} = g_{\pm,b}\(t,x=L\)\, W_t^{({\rm vac.)}}\,
g_{\pm,b}^{-1}\(t, x=-L\).
\ee
Using the definition of $g_{\pm,b}$ in \rf{gpmdef}, one gets that
\be
W_t^{(\pm)} \, \Psi_{2n+1}\, {W_t^{(\pm)}}^{-1} = e^{\pm \(2n+1\)\, C\,
  \(\xi_{2n+1}^{(\pm)}\(t,x=L\)-\xi_{2n+1}^{(\pm)}\(t,x=-L\)\)}\;
\Psi_{2n+1}.  
\ee
Thus, we have two infinite sets of conserved charges given by  
\be
\Omega_{2n+1}^{(\pm)}\equiv\pm \(2n+1\)\,\(
\xi_{2n+1}^{(\pm)}\(t,x=L\)-\xi_{2n+1}^{(\pm)}\(t,x=-L\)\) 
 \qquad \qquad \quad n=0,1,2,\ldots
\lab{chargesxi}
\ee
In order to evaluate those charges we use the expressions for the
solutions, on the orbit of the vacuum, given by the dressing
transformation method. Using \rf{gaussdecomp}, \rf{gpmdef},
\rf{taudef1}, \rf{g0eigenvalue}, and the 
results of the appendix \ref{sec:sl2km} we find that (for $n\geq 0$, and
$i=0,1$)  
\br
 \bra{\lambda_i}
\Psi_{\rm vac}\,h\,\Psi_{\rm vac}^{-1}\, b_{-2n-1}\ket{\lambda_i}&\equiv& 
= \bra{\lambda_i}G_{-}^{-1}\, G_0^{-1}\, G_{+}\,
b_{-2n-1}\ket{\lambda_i}
\nonu\\
&=&\tau_i\; 
\bra{\lambda_i}g_{+,F}^{-1}\,g_{+,b}\,
b_{-2n-1}\,g_{+,b}^{-1}\ket{\lambda_i}
\nonu\\
&=&\tau_i\; 
\bra{\lambda_i}g_{+,F}^{-1}\(b_{-2n-1}+(2n+1)\, \xi^{(+)}_{2n+1}\,
C\)\ket{\lambda_i} 
\nonu\\
&=&\tau_i\;\( 
\bra{\lambda_i}g_{+,F}^{-1}\, b_{-2n-1}\ket{\lambda_i} 
+(2n+1)\, \xi^{(+)}_{2n+1}\).\nonu
\er 
Now using \rf{bcgpmf} we get
\be
\xi^{(+)}_{2n+1}\(t,x=\pm L\) = \frac{1}{\(2n+1\)}\; 
\frac{\bra{\lambda_i}
\Psi_{\rm vac}\,h\,\Psi_{\rm vac}^{-1}\,
b_{-2n-1}\ket{\lambda_i}}{\bra{\lambda_i} 
\Psi_{\rm vac}\,h\,\Psi_{\rm vac}^{-1}\ket{\lambda_i}}\mid_{x=\pm L}.
\lab{xiplussg}
\ee
Using similar arguments we also see that
\be
\xi^{(-)}_{2n+1}\(t,x=\pm L\) =-\; \frac{1}{\(2n+1\)}\; 
\frac{\bra{\lambda_i}b_{2n+1}\,
\Psi_{\rm vac}\,h\,\Psi_{\rm vac}^{-1}
\ket{\lambda_i}}{\bra{\lambda_i} 
\Psi_{\rm vac}\,h\,\Psi_{\rm vac}^{-1}\ket{\lambda_i}}\mid_{x=\pm L}.
\lab{ximinussg}
\ee
Therefore, the charges \rf{chargesxi} become 
\be
\Omega_{2n+1}^{(+)}=\frac{\bra{\lambda_i}
\Psi_{\rm vac}\,h\,\Psi_{\rm vac}^{-1}\,
b_{-2n-1}\ket{\lambda_i}}{\bra{\lambda_i} 
\Psi_{\rm vac}\,h\,\Psi_{\rm
  vac}^{-1}\ket{\lambda_i}}\mid_{x=-L}^{x=+L}
\lab{chargesplus}
\ee
and
\be
\Omega_{2n+1}^{(-)}=\frac{\bra{\lambda_i}b_{2n+1}\,
\Psi_{\rm vac}\,h\,\Psi_{\rm vac}^{-1}
\ket{\lambda_i}}{\bra{\lambda_i} 
\Psi_{\rm vac}\,h\,\Psi_{\rm
  vac}^{-1}\ket{\lambda_i}}\mid_{x=-L}^{x=+L}.
\lab{chargesminus}
\ee
In particular, using \rf{importantrel1}, \rf{energy-momentum-formula},
and \rf{chargesxi} we see that the energy and momentum of the
solutions on the orbit of the vacuum are given, respectively, by:
\be
E = 2\(\Omega_{1}^{(+)}-\Omega_{1}^{(-)}\), \qquad \qquad \qquad
P= -2\(\Omega_{1}^{(+)}+\Omega_{1}^{(-)}\).
\lab{ep}
\ee

\subsection{Soliton solutions}

The soliton solutions are not only the most important ones in the
orbit of the vacuum, but also the simplest ones to construct using the
dressing method. The $n$-soliton solutions are obtained by taking the
constant group element $h$ introduced in \rf{gaussdecomp} as the
product of $n$ 
exponentials of eigenvectors of the oscillators $b_{2n+1}$
\cite{olivetensor,olivesolitonicspec,fmg}, namely the 
vertex operators defined in \rf{vertexdef}, {\it i.e.}
\be
h=\prod_{i=1}^n e^{a_i\,V\(z_i\)}.
\lab{solitonh}
\ee
Therefore, using \rf{psivac} and \rf{vertexeigen} one gets that
\be
\Psi_{{\rm vac}}\, h\, \Psi_{{\rm vac}}^{-1} = 
\prod_{i=1}^n e^{a_i\,e^{\Gamma\(z_i\)}\,V\(z_i\)}= 
\prod_{i=1}^n \(1+a_i\,e^{\Gamma\(z_i\)}\,V\(z_i\)\),
\lab{nsolitontauoperator}
\ee
where we have used the nilpotency property \rf{vertexnipotent} of the
vertex operator and have introduced
\be
\Gamma\(z_i\)\equiv z_i\, x_{+}-\frac{x_{-}}{z_i}. 
\lab{Gammadef}
\ee
Using \rf{vertexproductn} and \rf{vertexvev}, one then gets that the
tau-functions \rf{taudef1} are given by
\br
\tau_j &=& 1 + (-1)^j \sum_{l=1}^n a_l \, e^{\Gamma\(z_l\)} + 
 \sum_{l_1<l_2=1}^n 
\(\frac{z_{l_1}-z_{l_2}}{z_{l_1}+z_{l_2}}\)^2\, 
 a_{l_1}\, a_{l_2} \, e^{\Gamma\(z_{l_1}\)+\Gamma\(z_{l_2}\)}+\ldots
\nonu\\
&+&(-1)^j\sum_{l_1<l_2<l_3=1}^n 
\(\frac{z_{l_1}-z_{l_2}}{z_{l_1}+z_{l_2}}\)^2 
\(\frac{z_{l_1}-z_{l_3}}{z_{l_1}+z_{l_3}}\)^2
\(\frac{z_{l_2}-z_{l_3}}{z_{l_2}+z_{l_3}}\)^2 
a_{l_1} a_{l_2}  a_{l_3}
e^{\Gamma\(z_{l_1}\)+\Gamma\(z_{l_2}\)+\Gamma\(z_{l_3}\)} 
\nonu\\
&& \ldots + (-1)^{j\,n} 
\prod_{k_1<k_2=1}^{n}
\(\frac{z_{k_1}-z_{k_2}}{z_{k_1}+z_{k_2}}\)^2 
\prod_{l=1}^n a_l \,e^{\Gamma\(z_{l}\)}
\qquad\qquad {\rm for}
\quad j=0,1.
\lab{nsolitonsolution}
\er
The solution for the fields are then obtained through \rf{taudef2}. 

Observe that using \rf{vertexeigen} one has that
$$
b_{2n+1}\(1+a \,e^{\Gamma\(z \)}\,V\(z \)\) =
\(1+a \,e^{\Gamma\(z \)}\,V\(z \)\)b_{2n+1} 
- 2\, z ^{2n+1}a \,e^{\Gamma\(z \)}\,V\(z \).
$$
Thus, using \rf{nsolitontauoperator}, the charges
\rf{chargesplus} and \rf{chargesminus} for the 
$n$-soliton sector of solutions,  become ($l=0,1$) 
\br
&&\Omega_{2n+1}^{(\pm)}= \pm 2\,\sum_{k=1}^n z_k^{\mp\(2n+1\)}a_k
\,e^{\Gamma\(z_k\)}\,
\times 
\lab{chargesnsoliton1}\\
&&\times
\frac{\bra{\lambda_l}
\left[\prod_{i=1}^{k-1} \(1+a_i\,e^{\Gamma\(z_i\)}\,V\(z_i\)\)\right]
V\(z_k \)
\left[\prod_{j=k+1}^{n} \(1+a_j\,e^{\Gamma\(z_j\)}\,V\(z_j\)\)\right]
\ket{\lambda_l}}{\bra{\lambda_l}
\prod_{i=1}^n \(1+a_i\,e^{\Gamma\(z_i\)}\,V\(z_i\)\) 
\ket{\lambda_l}}\mid_{x=-L}^{x=+L}.
\nonu
\er
Let us now parametrize $z_i$ as
\be
z_i = e^{-\alpha_i + i \theta_i} \qquad \qquad\qquad \qquad 
\mbox{\rm with $\alpha_i$ and $\theta_i$ real.}
\ee
Then \rf{Gammadef} becomes
\be
\Gamma\(z_i\)=\frac{1}{\sqrt{1-v_i^2}}\,\left[
  \cos\theta_i\(x-v_i\,t\)+i\,\sin\theta_i\,\(t-v_i\,x\)\right], 
\lab{finalgamma}
\ee
where $v_i$ are velocities in units of the speed of light and
\be
v_i=\tanh \alpha_i,\qquad \qquad\qquad \cosh\alpha_i=\frac{1}{\sqrt{1-v_i^2}}.
\ee

Note that the behaviour of $e^{\Gamma\(z_i\)}$ as $x\rightarrow \pm
  \infty$ is determined by the sign of $\cos\theta_i$. In addition, if
  a given combination of exponentials of $\Gamma$'s dominates the
  behaviour of the 
  denominator of a given term of the sum in \rf{chargesnsoliton1} for
  $x\rightarrow \pm \infty$  
  then the same combination dominates the behaviour of the numerator
  of that term. Consequently, the corresponding
  expectation value of the product of $V$'s cancels out and we have that 
\be
\Omega_{2n+1}^{(\pm)}= \pm 2\,\sum_{k=1}^n \epsilon_k\,
z_k^{\mp\(2n+1\)}, 
\lab{chargesnsoliton2}
\ee
where $\epsilon_k=\pm 1$ are signs determined by the dominant
combinations of exponentials of $\Gamma$s. In consequence, the form of
  the charges 
for the $n$-soliton sector of solutions is quite simple.

\subsubsection{$1$-soliton sector}

In order to have a real solution for the field $\vp$ in  the
$1$-soliton sector we need to take either $\theta=0$ (soliton) or
$\theta=\pi$ (anti-soliton). In addition, we need $a$ to be pure
imaginary. Then from \rf{nsolitonsolution} and
\rf{finalgamma} we have $\tau_0=\tau_1^*$ and from \rf{taudef2}
\be
\vp = 4\, {\rm
  ArcTan}\left[\exp\(\varepsilon\frac{\(x-v\,t-x_0\)}{\sqrt{1-v^2}}\)\right],
\ee
where we have taken $a=i\exp\(-\varepsilon \, x_0/\sqrt{1-v^2}\)$, and
$\varepsilon\equiv e^{i\theta}=\pm 1$, $\theta=0,\pi$. 

Evaluating the charges \rf{chargesnsoliton1} one gets
\be
\Omega_{2n+1}^{(\pm)}= \pm 2\,
\varepsilon\,z^{\mp\(2n+1\)} =\pm 2\, e^{\pm\(2n+1\)\alpha}=
\pm 2\,\left[\frac{1+v}{1-v}\right]^{\pm\frac{\(2n+1\)}{2}}.
\ee
In particular, the energy and momentum \rf{ep} become
\be
E=\frac{8}{\sqrt{1-v^2}},\qquad\qquad\qquad P=-\frac{8\,v}{\sqrt{1-v^2}}.
\ee

\subsubsection{$2$-soliton sector}

In this sector we have two types of real solutions: unbounded 
 $2$-solitons and breathers. 

\noindent{\bf Unbounded $2$-soliton solutions}

In this case we take $\theta_1 , \theta_2 = 0, \pi$, corresponding to
the choices of solitons or anti-solitons, and also take $a_i$, $i=1,2$
pure imaginary. We then have $\tau_0=\tau_1^*$ and
\be
\tau_0 = 1 + i\, e^{\Gamma_1}+i\, e^{\Gamma_2}-
\(\frac{1-\varepsilon_1\,\varepsilon_2\, e^{\alpha_1-\alpha_2}}{
1+\varepsilon_1\,\varepsilon_2\, e^{\alpha_1-\alpha_2}}\)^2\, 
e^{\Gamma_1+\Gamma_2}
\ee
with
\be
\Gamma_i =
\varepsilon_i\,\frac{\(x-v_i\,t-x_0^{(i)}\)}{\sqrt{1-v_i^2}}
\ee
with $a_i=i\exp\(-\varepsilon_i \, x_0^{(i)}/\sqrt{1-v_i^2}\)$,
$\varepsilon_i\equiv e^{i\theta_i}=\pm 1$, $\theta_i=0,\pi$, $i=1,2$.

Evaluating the charges \rf{chargesnsoliton1} we get
\br
\Omega_{2n+1}^{(\pm)}&=& \pm 2\,\(
\varepsilon_1\,z_1^{\mp\(2n+1\)}+\varepsilon_2\,z_2^{\mp\(2n+1\)} \)
=\pm 2\, \(e^{\pm\(2n+1\)\alpha_1}+e^{\pm\(2n+1\)\alpha_2}\)\nonu\\
&=&
\pm 2\,\( \left[\frac{1+v_1}{1-v_1}\right]^{\pm\frac{\(2n+1\)}{2}}+
\left[\frac{1+v_2}{1-v_2}\right]^{\pm\frac{\(2n+1\)}{2}}\).
\er
Therefore 
\be
E=\frac{8}{\sqrt{1-v_1^2}}+\frac{8}{\sqrt{1-v_2^2}},
\qquad\qquad\qquad 
P=-\frac{8\,v_1}{\sqrt{1-v_1^2}}-\frac{8\,v_2}{\sqrt{1-v_2^2}}.
\ee

\noindent{\bf Breathers}

For the breather solutions we take $\alpha_1=\alpha_2\equiv\alpha$,
$\theta_1=-\theta_2\equiv \theta$, and so $z_1=z_2^*$. We also
take $a_1=-a_2=-{\rm cotan}\, \theta$. We then have
$\Gamma\(z_1\)=\Gamma\(z_2\)^*$ and, again, $\tau_0=\tau_1^*$, with
\be
\tau_0 = 1+e^{2\,\Gamma_R} + 2\,i\,\({\rm cotan}\, \theta\)\;
e^{\Gamma_R}\, \sin \Gamma_I,
\ee
where
\be
\Gamma_R = \frac{\cos\theta}{\sqrt{1-v^2}}\, \(x-v\,t\),\qquad\qquad 
\Gamma_I = \frac{\sin\theta}{\sqrt{1-v^2}}\, \(t-v\,x\).
\ee
Therefore
\be
\vp = 4\, {\rm Arctan}\frac{\({\rm cotan}\, \theta\)\,\sin
  \Gamma_I}{\cosh \Gamma_R}.
\ee
Evaluating the charges \rf{chargesnsoliton1} one gets
\br
\Omega_{2n+1}^{(\pm)}&=& \pm 2\,\varepsilon\, \(
e^{\mp\(2n+1\)\(-\alpha+i\theta\)}
+e^{\mp\(2n+1\)\(-\alpha-i\theta\)}\)
\nonu\\
&=&\pm
4\,\varepsilon\,\left[\frac{1+v}{1-v}\right]^{\pm\frac{\(2n+1\)}{2}}\,
\cos\left[\(2n+1\)\,\theta\right],
\er
where $\varepsilon \equiv {\rm sign}\(\cos\theta\)$. Therefore, the
energy and momentum become
\be
E= \frac{16\,\mid\cos\theta\mid}{\sqrt{1-v^2}}, \qquad \qquad \qquad \qquad 
P= -\frac{16\,v\,\mid\cos\theta\mid}{\sqrt{1-v^2}}.
\ee

\subsubsection{$N$-soliton sector}

As shown in \rf{chargesnsoliton2}, the conserved charges evaluated on
the solutions coming from the choice \rf{solitonh} for the constant
group element $h$ of the dressing method have an additive
character. Therefore, if one considers a solution with $N$ solitons
and $M$ breathers the charges are given by
\be
\Omega_{2n+1}^{(\pm)} = \pm 2\,\sum_{i=1}^N 
\left[\frac{1+v_i}{1-v_i}\right]^{\pm\frac{\(2n+1\)}{2}} 
\pm 4\,\sum_{j=1}^M
\varepsilon_j\,\left[\frac{1+v_j}{1-v_j}\right]^{\pm\frac{\(2n+1\)}{2}}\,
\cos\left[\(2n+1\)\,\theta_j\right]
\ee
with $\varepsilon_j={\rm sign}\(\cos\theta_j\)$. Consequently the
energy and momentum are also additive and one has
\br
E&=&\sum_{i=1}^N \frac{8}{\sqrt{1-v_i^2}}+\sum_{j=1}^M
\frac{16\,\mid\cos\theta_j\mid}{\sqrt{1-v_j^2}},
\nonu\\
P&=&-\sum_{i=1}^N \frac{8\,v_i}{\sqrt{1-v_i^2}}-\sum_{j=1}^M
\frac{16\,v_j\,\mid\cos\theta_j\mid}{\sqrt{1-v_j^2}}.
\er

\section{Generalized soliton hierarchies}
\label{sec:general}
\setcounter{equation}{0}

The results obtained above for the sine-Gordon model can certainly be
generalized to other theories possessing soliton solutions. We sketch
here how this can be done using the basic structures known to be
responsible for 
the existence of solitons.  As explained for instance in \cite{fmg},  
practically all two dimensional exact soliton solutions known in the 
literature belong to a class that can be characterized by the
following features: 
\begin{enumerate} 
\item They are solutions of two dimensional theories that admit a zero
  curvature representation of their equations of motion, {\it i.e.} there
  exist potentials (Lax operators) $A_{\mu}$, which are functionals of
  the fields of the theory and which belong to a Kac-Moody
  algebra ${\cal G}$ such that the condition
\be
\sbr{\partial_{\mu} + A_{\mu}}{\partial_{\nu} + A_{\nu}} =0
\lab{zcgen}
\ee
is equivalent to the equations of motion. The indices $\mu ,\nu$
  correspond to the two coordinates of space-time, or to the  various
  times $t_N$ 
  of a hierarchy of soliton theories (see \cite{fmg} for details). 
\item There exist an integer gradation of ${\cal G}$ 
\be
{\cal G} = \oplus_{n\in \IZ} {\cal G}_n, \qquad \qquad 
\sbr{{\cal G}_m}{{\cal G}_n} \subset {\cal G}_{m+n}
\lab{gradation}
\ee
such that the potentials can be decomposed as
\be
A_{\mu}=\sum_{n=N_{\mu}^{-}}^{N_{\mu}^{+}} A_{\mu}^{(n)}, \qquad \qquad
{\rm where} \qquad A_{\mu}^{(n)} \in {\cal G}_n
\lab{gradesamu}
\ee
with $N_{\mu}^{-}$ and $N_{\mu}^{+}$ being non-positive and
non-negative integers, respectively.
\item There exist at least one ``vacuum solution'' of the theory such
  that the potentials $A_{\mu}$ evaluated on it  belong to an abelian
  subalgebra, up to central term, of ${\cal G}$, {\it i.e.}
\be
A_{\mu}^{({\rm vac})} = \sum_{n=N_{\mu}^{-}}^{N_{\mu}^{+}} \sum_{a=1}^r
c_{\mu}^{a,n}\, b_{n}^a + \sigma_{\mu}\, C 
\equiv E_{\mu}  + \sigma_{\mu}\, C,
\lab{avacgen}
\ee
where $c_{\mu}^{a,n}$ are constants, $C$ is the central element of
${\cal G}$, and $b_{n}^a$ satisfy 
an algebra of oscillators (Heisenberg subalgebra) 
\be
\sbr{b_{m}^a}{b_{n}^b}= \omega^{ab}\, m\, \delta_{m+n,0}\, C
\lab{heisenbergalg}
\ee
with $\omega^{ab}$ being a symmetric matrix, and $a,b=1,2,\ldots r$, 
labels the number 
of infinite sets of oscillators. The index $n$ corresponds to the grade
of the oscillators, {\it i.e.} $b_{m}^a\in {\cal G}_n$, and they do not have
to exist for all values of $n$ (for instance,  in the case
of sine-Gordon, as discussed in section \ref{sec:sg} and appendix
\ref{sec:sl2km}, they exist only for odd $n$) 
\end{enumerate}

The soliton solutions are then constructed using the dressing method
in a manner similar to that explained in section \ref{sec:sg} for the
sine-Gordon model. Since $A_{\mu}^{({\rm vac})}$, given in
\rf{avacgen}, satisfy the zero 
curvature equation \rf{zcgen}, 
there exists a group element $\Psi_{{\rm vac}}$, which is an
exponentiation of $E_{\mu}$ (oscillators) and $C$ (see \cite{fmg} for
details), such that
$$
A_{\mu}^{({\rm vac})} = - \partial_{\mu}\Psi_{{\rm vac}}\,
\Psi_{{\rm vac}}^{-1} .
$$
We then choose a constant group element $h$ such that there exists a 
Gauss like decomposition
\be
\Psi_{{\rm vac}}\, h\, \Psi_{{\rm vac}}^{-1} =
G_{-}^{-1}\,G_{0}^{-1}\,G_{+}
\lab{hdefgen}
\ee
with $G_{+,0,-}$ being group elements obtained by the exponentiation of
generators of ${\cal G}$ with 
positive, zero, and negative grades, respectively, with respect to
\rf{gradation}. We now introduce
\be
\Psi_h \equiv G_0\,G_{-}\,\Psi_{{\rm vac}}\, h = G_{+}\, \Psi_{{\rm vac}},
\qquad \qquad 
{\bar \Psi}_h \equiv  G_{-}\,\Psi_{{\rm vac}}\, h = 
G_0^{-1}\,G_{+}\, \Psi_{{\rm vac}}
\lab{psihgendef}
\ee
and the corresponding potentials
\be
A_{\mu}^{h} \equiv - \partial_{\mu}\Psi_{h}\,\Psi_{h}^{-1}, \qquad
\qquad \qquad  
{\bar A}_{\mu}^{h} \equiv - \partial_{\mu}{\bar \Psi}_{h}\,{\bar
  \Psi}_{h}^{-1}.
\lab{ahgendef}
\ee
Therefore one has 
\br
A_{\mu}^{h} &=& G_{+}\,A_{\mu}^{({\rm vac})}\, G_{+}^{-1}-
\partial_{\mu}G_{+}\, G_{+}^{-1}
\lab{dressinggen1}\\
&=& G_{0}\, G_{-}\,A_{\mu}^{({\rm vac})}\, \(G_{0}\, G_{-}\)^{-1}-
\partial_{\mu}\(G_{0}\, G_{-}\)\, \(G_{0}\, G_{-}\)^{-1},
\lab{dressinggen2}\\
{\bar A}_{\mu}^{h} &=& G_{-}\,A_{\mu}^{({\rm vac})}\, G_{-}^{-1}-
\partial_{\mu}G_{-}\, G_{-}^{-1}
\lab{dressinggen3}\\
&=& G_{0}^{-1}\, G_{+}\,A_{\mu}^{({\rm vac})}\, \(G_{0}^{-1}\, G_{+}\)^{-1}-
\partial_{\mu}\(G_{0}^{-1}\, G_{+}\)\, \(G_{0}^{-1}\, G_{+}\)^{-1}.
\lab{dressinggen4}
\er 

Using arguments similar to those given before \rf{dress1}-\rf{dress2} one can
then show that $A_{\mu}^{h}$ and ${\bar A}_{\mu}^{h}$ have the same grading 
structure as $A_{\mu}$ in \rf{gradesamu}. Indeed, using \rf{avacgen}
one sees that \rf{dressinggen1} implies that $A_{\mu}^{h}$ has
components of grade greater or equal to $N_{\mu}^{-}$ and
\rf{dressinggen2} implies that it has components of grade smaller or
equal to $N_{\mu}^{+}$. Therefore,  $A_{\mu}^{h}$ must have components
of grade varying from $N_{\mu}^{-}$ to $N_{\mu}^{+}$. Using similar
arguments for \rf{dressinggen3} and \rf{dressinggen4} one concludes
that ${\bar A}_{\mu}^{h}$  must also have components
of grade varying from $N_{\mu}^{-}$ to $N_{\mu}^{+}$. In fact, from their
definition \rf{ahgendef}, one notices that
$A_{\mu}^{h}$ and ${\bar A}_{\mu}^{h}$ are related by  a gauge
transformation with the group element $G_0$, which involves only zero
grade generators, and so they must indeed have the same grading
structure. Thus, $A_{\mu}^{h}$ corresponds to $A_{\mu}$ evaluated
on the solution constructed by the dressing method. By equating
$A_{\mu}^{h}$ to $A_{\mu}$ given in \rf{gradesamu}, which is 
a functional of the underlying fields, one then defines the solution
of their equations of motion. Note that we could as well have equated
 ${\bar A}_{\mu}^{h}$  to $A_{\mu}$,
as this corresponds to a gauge choice which we can make in order 
to get the relation among the parameters of $G_{0,\pm}$ with the
fields as simple as possible.  

The soliton solutions are obtained by choosing the constant group
element $h$, introduced in \rf{hdefgen}, as
\be
h= \prod_{l=1}^N\prod_{k=1}^n e^{a_{l,k}\, V_l\(z_k\)},
\lab{hforsolitongen}
\ee
where $ V_l\(z_k\)$ are eigenvectors of the operators $E_{\mu}$, introduced
in \rf{avacgen},
$$
\sbr{E_{\mu}}{V_l\(z_k\)}= \lambda_{\mu}^l\(z_k\) \, V_l\(z_k\)
$$
and where $l$ labels the species or types of solitons and $z_k$ are
parameters that determine the velocities and topological charges of the
solitons (see \cite{fmg} for more details).  In the case of
the sine-Gordon model discussed in section \ref{sec:sg}, we have seen that
there exist only one species of solitons. 

Denoting by ${\cal H}$ the (Heisenberg) subalgebra generated by the
oscillators $b_n^a$'s and $C$, and by ${\cal F}$ its complement in
the Kac-Moody algebra ${\cal G}$ we see that 
\be
{\cal G} = {\cal H} + {\cal F}.
\lab{decompgen}
\ee

Then we split the group elements $G_{\pm}$ according to such a
decomposition, {\it i.e.}
\be
G_{\pm}= g_{\pm ,F}^{-1}\, g_{\pm ,b}
\lab{gpmdecompgen}
\ee
with
\be
g_{\pm ,b} = \exp\(\sum_{n=1}^{\infty}\sum_{a=1}^r \xi_{a,n}^{\(\pm\)}\,
  b_{\pm n}^a \)
\lab{gbgendef}
\ee
and $g_{\pm ,F}$ being group elements obtained by exponentiating the
parts ${\cal F}_{+}$ and ${\cal F}_{-}$  of ${\cal F}$
containing the generators of positive and negative grades respectively,
{\it i.e.} $g_{\pm ,F}= \exp\({\cal F}_{\pm}\)$. In this case the relations
\rf{dressinggen1} and \rf{dressinggen3} can be rewritten, respectively, as 
\br
g_{+,F}\,A_{\mu}^h\,g_{+,F}^{-1}-\partial_{\mu}g_{+,F}\,g_{+,F}^{-1} &=& 
g_{+,b}\,A^{({\rm
    vac.})}_{\mu}\,g_{+,b}^{-1}-\partial_{\mu}g_{+,b}\,g_{+,b}^{-1}\equiv 
a_{\mu}^{(+)},  
\lab{nicegenrel1}
\\
g_{-,F}\,{\bar
  A}_{\mu}^h\,g_{-,F}^{-1}-\partial_{\mu}g_{-,F}\,g_{-,F}^{-1} &=&  
g_{-,b}\,A^{({\rm
    vac.})}_{\mu}\,g_{-,b}^{-1}-\partial_{\mu}g_{-,b}\,g_{-,b}^{-1}\equiv 
a_{\mu}^{(-)},  
\lab{nicegenrel2}
\er
where we have introduced the potentials $a_{\mu}^{(\pm)}$. 

The conserved charges can now be constructed in a manner similar to
that of section \ref{sec:sg} of the sine-Gordon case. Denote by $x$
and $t$ the space-time coordinates for our generalized soliton
theory. Suppose that the time component of the potential introduced
in \rf{gradesamu} satisfies the boundary condition \rf{bc}. Then  using
\rf{nicegenrel1}-\rf{nicegenrel2} and the
arguments leading to \rf{wtnew} one sees that
\be
W_t^{(\pm)} = P\, e^{-\int_{x=-L}^{x=L}dx\, a_x^{(\pm)}} = 
g_{\pm,b}\(t,x=L\)\, W_t^{({\rm vac.)}}\,
g_{\pm,b}^{-1}\(t, x=-L\) ,
\lab{wtgen}
\ee
where
\be
W_t^{({\rm vac.)}}= P\, e^{-\int_{x=-L}^{x=L}dx\, A_{x}^{({\rm vac})}}.
\ee
For every pair of oscillators, $b_n^a$ and $b_{-n}^a$, we introduce the
operator ($n>0$)
\be
\Psi_{a,n} = \; : e^{ b_{n}^a+ b_{-n}^a}:\;\;  =  
e^{ b_{-n}^a}\; e^{ b_{n}^a} 
\lab{psidefgen}
\ee
where $::$, as before, denotes the normal ordering of the oscillators,
{\it i.e.} positive 
grade oscillators are put to the right of the negative ones. 
Then, using \rf{heisenbergalg} and \rf{gbgendef} we see that
\be
W_t^{(\pm)}\, \Psi_{a,n}\, {W_t^{(\pm)}}^{-1} =
e^{\(\Omega^{(\pm)}_{a,n}+\Omega^{({\rm vac})}_{a,n}\)\, C}\, \Psi_{a,n}
\ee
with 
\be
\Omega^{(\pm)}_{a,n} = \pm \, n\, \sum_{b=1}^r \omega^{ab}\( 
\xi_{b,n}^{\(\pm\)}\(t,x=L\)-\xi_{b,n}^{\(\pm\)}\(t,x=-L\)\)
\lab{chargesgen}
\ee
and $\Omega^{({\rm vac})}_{a,n}$ are the vacuum values of the charges
\be
W_t^{({\rm vac.)}}\, \Psi_{a,n}\,{W_t^{({\rm vac.)}}}^{-1} =
e^{\Omega^{({\rm vac})}_{a,n}\, C}\, \Psi_{a,n}.
\lab{wtvacgen}
\ee

In a manner similar to that of the sine-Gordon case, the parameters
$\xi_{a,n}^{\(\pm\)}$ are determined for the solution associated to a
given constant group element $h$, by the matrix elements  of the form 
$\bra{\lambda}\Psi_{{\rm vac}}\, h\, \Psi_{{\rm vac}}^{-1}\,b_n^a
\ket{\lambda}$   and 
$\bra{\lambda}\,b_{-n}^a\,\Psi_{{\rm vac}}\, h\, \Psi_{{\rm vac}}^{-1}
\ket{\lambda}$ of the
operators in \rf{hdefgen}, with $n>0$, and $\ket{\lambda}$ being a
highest weight 
state of a given representation of the Kac-Moody algebra ${\cal G}$. 

Note that, using \rf{avacgen} and \rf{gbgendef}, the r.h.s. of
\rf{nicegenrel1} and \rf{nicegenrel2} give  
\br
a_{\mu}^{(+)}&=&  E_{\mu}  + \sigma_{\mu}\, C +
\sum_{n=1}^{-N_{\mu}^{-}}\sum_{a,b=1}^r \omega^{ab}\, c_{\mu}^{a,-n}\, n\,
\xi_{b,n}^{\(+\)}\, C - 
\sum_{n=1}^{\infty}\sum_{a=1}^r \partial_{\mu} \xi_{a,n}^{\(+\)}\,
  b_{n}^a,
\nonu\\
a_{\mu}^{(-)}&=&  E_{\mu}  + \sigma_{\mu}\, C +
\sum_{n=1}^{N_{\mu}^{+}}\sum_{a,b=1}^r \omega^{ab}\, c_{\mu}^{a,n}\, (-n)\,
\xi_{b,n}^{\(-\)}\, C - 
\sum_{n=1}^{\infty}\sum_{a=1}^r \partial_{\mu} \xi_{a,n}^{\(-\)}\,
  b_{-n}^a .
\nonu
\er
Equating to the l.h.s. of \rf{nicegenrel1} and \rf{nicegenrel2} one
gets that the $C$-part gives
\br
\sum_{n=1}^{-N_{\mu}^{-}}\sum_{a,b=1}^r \omega^{ab}\, c_{\mu}^{a,-n}\, n\,
\xi_{b,n}^{\(+\)} &=& 
\(g_{+,F}\,A_{\mu}^h\,g_{+,F}^{-1}\)_{{\rm coeff.}\, C} - \sigma_{\mu},
\nonu\\
\sum_{n=1}^{N_{\mu}^{+}}\sum_{a,b=1}^r \omega^{ab}\, c_{\mu}^{a,n}\, (-n)\,
\xi_{b,n}^{\(-\)} &=& 
\(g_{-,F}\,{\bar A}_{\mu}^h\,g_{-,F}^{-1}\)_{{\rm coeff.}\, C} -
\sigma_{\mu}.
\nonu
\er
Therefore, from \rf{chargesgen} one gets
\br
\sum_{n=1}^{-N_{\mu}^{-}}\sum_{a,b=1}^r
c_{\mu}^{a,-n}\,\Omega^{(+)}_{a,n}  &=& 
\left[\(g_{+,F}\,A_{\mu}^h\,g_{+,F}^{-1}\)_{{\rm coeff.}\, C} - 
\sigma_{\mu}\right]_{x=-L}^{x=L},
\nonu\\
\sum_{n=1}^{N_{\mu}^{+}}\sum_{a,b=1}^r
c_{\mu}^{a,n}\,\Omega^{(-)}_{a,n}  
&=& 
\left[\(g_{-,F}\,{\bar A}_{\mu}^h\,g_{-,F}^{-1}\)_{{\rm coeff.}\, C} -
\sigma_{\mu}\right]_{x=-L}^{x=L}.
\er
If $\left[\(g_{+,F}\,A_{\mu}^h\,g_{+,F}^{-1}\)_{{\rm coeff.}\, C} - 
\sigma_{\mu}\right]$ and $\left[\(g_{-,F}\,{\bar
    A}_{\mu}^h\,g_{-,F}^{-1}\)_{{\rm 
    coeff.}\, C} - \sigma_{\mu}\right]$, can be expressed locally in
terms of the underlying 
fields of the theory, we see that the above linear combinations of
conserved charges are boundary terms. This happens for instance, in
the abelian and non-abelian Toda models \cite{acfgzhirota,fmgcat},
where the combinations of 
charges turn out to be related to the energy and momentum of the solutions.

\subsection{The example of the mKdV equation}
\label{sec:mkdv}

The modified Korteweg-de Vries equation (mKdV) is an example of a
soliton theory that fulfills the requirements described at the beginning of
section \ref{sec:general}, and so can have its conserved charges
calculated as described in this paper. As we pointed out it is
important to work with a zero curvature representation based on the
Kac-Moody algebra with a non vanishing central term. We use here the
zero curvature potentials for the mKdV equation constructed in section
IV.A of reference  \cite{fmg}. The potentials are given by 
\br
A_x&=& -b_1-q\, F_0-\nu\, C 
\lab{mkdvpot}\\
A_t&=& -b_3 - q\, F_2+\frac{1}{2} \, \partial_x q \, F_1+\frac{1}{2}\,
q^2\, b_1-\frac{1}{2}\,\(\frac{1}{2}\,\partial_x^2 q -q^3 \)\,
F_0-\frac{1}{16}\partial_x q^{2} \, C 
\nonu
\er
where $C$, $b_{j}$, $j=1,3$, and $F_k$, $k=0,1,2$, are generators of
the $sl(2)$ Kac-Moody algebra defined in appendix \ref{sec:sl2km}, and
which commutation relations are given in \rf{newbasiscomrel}. We have
denoted by $q$ the mKdV field, and by $\nu$ an extra field associated
to the central term $C$ of the algebra.  Replacing the
potentials \rf{mkdvpot} into the zero curvature condition  \rf{zcgen}
one gets that all components vanish with the exception of those in the
direction of $F_0$ and $C$ which give the equations of motion
\br
\partial_t q&=& \frac{1}{2}\,\partial_x\(\frac{1}{2}\,\partial_x^2 q-q^3 \)
\lab{mkdveq1}\\
\partial_t \nu&=& \frac{1}{16}\,\partial_x^2 q^{2}
\lab{mkdveq2}
\er
and \rf{mkdveq1} is the well known mKdV equation. 
Notice that the $\nu$ field is an expectant since it does affect the
equation of motion for the field $q$. That is similar to the $\rho$
field introduced in the sine-Gordon model in \rf{sgeq}. However, here
in the case of the mKdV equation we would not have to introduce such
field to work with a non-vanishing central term $C$. The reason is
that, contrary to the sine-Gordon case, all the generators appearing in
the potentials \rf{mkdvpot} have non negative grades w.r.t. the grading
operator $Q$ defined in \rf{graddef}. Therefore, the commutator term,
$\sbr{A_x}{A_t}$, of \rf{zcgen} does not produce  terms in the
direction of $C$. However as we show below, the introduction of such
field is important to make the dressing method consistent with a non
vanishing central term. In addition, that field is crucial for the
simple formula we obtain for the energy of the solutions. 
 
So, the mKdV theory fulfills the requirement $1$ at the beginning of
section \ref{sec:general}. As for the requirement $2$, we have that
the potentials \rf{mkdvpot} are decomposed as in \rf{gradesamu}
w.r.t. to the gradation defined by the grading operator $Q$ introduced
in \rf{graddef}. Indeed, one can check that in this case we have 
$N_x^{-}=0$, $N_x^+=1$, $N_t^-=0$ and $N_t^+=3$. The vacuum solution
of the requirement $3$ can be taken as $q=\nu=0$, and so the
potentials evaluated on it are given by
\be
A_x^{({\rm vac})} = -b_1 \qquad\qquad \qquad 
A_t^{({\rm vac})} = -b_3
\lab{vacpotmkdv}
\ee
Comparing with \rf{avacgen} we have $E_x=-b_1$, $E_t=-b_3$, and
$\sigma_{\mu}=0$. The relevant oscillator algebra \rf{heisenbergalg}
in this case is that generated by $b_{2n+1}$ (see
  \rf{newbasiscomrel}). The potentials \rf{vacpotmkdv} can be written
  as 
\be
A_{\mu}^{({\rm vac})} = -\partial_{\mu} \Psi_{\rm vac}\, \Psi_{\rm
  vac}^{-1} \qquad \quad {\rm with}\qquad \quad 
\Psi_{\rm vac}= e^{x\, b_1}\,e^{t\,b_3}
\ee
The dressing method can then be applied following the description
given from \rf{hdefgen} to \rf{dressinggen4}. With the vacuum
potential given by \rf{vacpotmkdv} it then follows from
\rf{dressinggen3}-\rf{dressinggen4} that ${\bar A}_{\mu}^h$ has the
same same grading structure as $A_{\mu}$ given in
\rf{mkdvpot}. We can then equate those two potentials in order to
evaluate the solutions. By comparing the 
zero grade part of ${\bar A}_{x}^h$ given in \rf{dressinggen4}, with
the zero grade part of $A_x$ given in \rf{mkdvpot}, one then gets that
\be
G_0=e^{\alpha \, F_0+\beta\,C} \qquad \quad {\rm with} \qquad \quad 
\partial_x\alpha=-q\; ;  \qquad \partial_x\beta=-\nu
\lab{g0mkdvdef}
\ee 
Since all the relations on the dressing method are valid on shell,
i.e. when the equations of motion hold true, we can use
\rf{mkdveq1}-\rf{mkdveq2} to get the time derivatives of the parameters
$\alpha$ and $\beta$. By taking the integration constants to vanish,
one obtains that 
\be
\partial_t \alpha= -
\frac{1}{2}\,\(\frac{1}{2}\,\partial_x^2 q-q^3 \)\; ; 
\qquad \qquad 
\partial_t \beta= -\frac{1}{16}\,\partial_x q^{2}
\lab{dtalphamkdv}
\ee
Replacing \rf{vacpotmkdv} into \rf{dressinggen1}-\rf{dressinggen2} one
observes that $A_{\mu}^h$ does not have the same grading structure as
$A_{\mu}$ given in \rf{mkdvpot}. In fact, contrary to $A_{\mu}$,
$A_{\mu}^h$ can not have zero grade components. In fact, $A_{\mu}^h$
corresponds to a potential ${\tilde A}_{\mu}$ obtained from $A_{\mu}$
of \rf{mkdvpot} by the gauge transformation (see
\rf{psihgendef}-\rf{ahgendef}) 
\be
{\tilde A}_{\mu}\equiv G_0\, A_{\mu}\, G_0^{-1}-\partial_{\mu} G_0\,
G_0^{-1}
\ee
Using \rf{g0mkdvdef} and \rf{dtalphamkdv} one gets
\br
 {\tilde A}_{x}&=& -\cosh\(2\,\alpha\)\, b_1-\sinh\(2\alpha\)\,F_1 
\lab{mkdvpot2}\\
{\tilde A}_t&=& -\cosh\(2\,\alpha\)\,b_3-\sinh\(2\alpha\)\,F_3 
- q\, F_2+\frac{1}{2} \, \left[\partial_x q \,
\cosh\(2\,\alpha\)+q^2\,\sinh\(2\alpha\)\right]\,F_1 \nonu\\
&+&\frac{1}{2}\, 
\left[ q^2\,\cosh\(2\,\alpha\)+\partial_x q \,\sinh\(2\alpha\)\right]\,b_1
\nonu
\er
So, ${\tilde A}_{\mu}$ is local in the parameter $\alpha$ but not on
the mKdV field $q$. In addition, it does not involve the extra filed
$\nu$ and neither the parameter $\beta$. Notice that, the vanishing of
the integration constants leading to  \rf{dtalphamkdv} is a
requirement of the dressing method, since if those constants were not
zero, 
${\tilde A}_{\mu}$ would have zero grade components. 

Using \rf{hdefgen} we now introduce the Hirota's tau functions
\br
\tau_0&=&\bra{\lambda_0}\Psi_{\rm vac}\,h\,\Psi_{\rm
  vac}^{-1}\ket{\lambda_0} = \bra{\lambda_0}G_0^{-1}\ket{\lambda_0}= 
e^{\frac{1}{2}\,\alpha-\beta}\nonu\\
\tau_1&=&\bra{\lambda_1}\Psi_{\rm vac}\,h\,\Psi_{\rm
  vac}^{-1}\ket{\lambda_1} = \bra{\lambda_1}G_0^{-1}\ket{\lambda_1}= 
e^{-\frac{1}{2}\,\alpha-\beta}
\lab{hirotataumkdv}
\er
where $\ket{\lambda_i}$, $i=0,1$, are the highest weight states of the
two fundamental representations of the $sl(2)$ Kac-Moody algebra, and
where 
we have used their properties given
in \rf{highestweight1}-\rf{highestweight2}. Therefore, using
\rf{g0mkdvdef} and \rf{hirotataumkdv}, 
the relation
among the fields and tau functions are given by 
\be
q=\partial_x\ln\frac{\tau_1}{\tau_0} \qquad \qquad \qquad \qquad 
\nu = \frac{1}{2}\, \partial_x\ln\(\tau_0\,\tau_1\)
\lab{fieldtaurelmkdv}
\ee
As explained in \rf{hforsolitongen} the soliton solutions, on the orbit
of the vacumm \rf{vacpotmkdv}, are obtained
by taking the constant group element $h$ to be exponentials of the
eigenvectors of $b_1$ and $b_3$. Evaluating the matrix elements in
\rf{hirotataumkdv} and replacing them into \rf{fieldtaurelmkdv} one
gets the solutions for the mKdV field $q$. 

The decomposition \rf{decompgen} in the case of the mKdV is such that
${\cal H}$ is generated by the oscillators $b_{2n+1}$,  and
the complement ${\cal F}$ by the generators $F_n$, with $n\in \IZ$. We
then write the group elements introduced in \rf{gpmdecompgen} as 
\be
g_{\pm ,b}=\exp\(\sum_{n=0}^{\infty} \xi_{2n+1}^{(\pm)} \, b_{\pm
  (2n+1)}\)\qquad \qquad 
g_{\pm ,F}=\exp\(\sum_{n=1}^{\infty} \zeta_n^{(\pm )} \, F_{\pm n}\)
\lab{gpmdefmkdv}
\ee 
The $x$-components of the relations \rf{nicegenrel1}-\rf{nicegenrel2}
are then given by  
\br
g_{+,F}\,{\tilde A}_{x}\,g_{+,F}^{-1}-\partial_{x}g_{+,F}\,g_{+,F}^{-1} &=& 
- g_{+,b}\,b_1\,g_{+,b}^{-1}-\partial_{x}g_{+,b}\,g_{+,b}^{-1}\equiv 
a_{x}^{(+)}  
\lab{nicemkdvrel1}
\\
&=& -b_1-\sum_{n=0}^{\infty}\partial_x \xi_{2n+1}^{(+)}\, b_{2n+1}
\nonu
\er
and
\br
g_{-,F}\,A_{x}\,g_{-,F}^{-1}-\partial_{x}g_{-,F}\,g_{-,F}^{-1} &=&  
-g_{-,b}\,b_1\,g_{-,b}^{-1}-\partial_{x}g_{-,b}\,g_{-,b}^{-1}\equiv 
a_{x}^{(-)} 
\lab{nicemkdvrel2}\\
&=& -b_1+\xi_1^{(-)}\, C - 
\sum_{n=0}^{\infty}\partial_x \xi_{2n+1}^{(-)}\, b_{-(2n+1)}
\nonu
\er
with $A_{x}$ and ${\tilde A}_{x}$ given by \rf{mkdvpot} and
\rf{mkdvpot2} respectively. The r.h.s. of \rf{nicemkdvrel1} and
\rf{nicemkdvrel2} do not contain terms in the direction of $F_n$. By
imposing the cancellation of the coefficients of $F_n$ on the l.h.s. of those
equations one determines  
the parameters $\zeta_n^{(\pm)}$. The first of them are given by 
\br
\zeta_1^{(-)}&=& -\frac{1}{2}\, q \qquad \qquad \qquad \,
\partial_x\zeta_1^{(+)}=-\sinh\(2\alpha\)
\lab{zetarelmkdv}\\
\zeta_2^{(-)}&=& \frac{1}{4}\, \partial_x q\qquad \qquad \qquad 
\partial_x\zeta_2^{(+)}=-2\,\zeta_1^{(+)}\, \cosh \(2\alpha\)
\nonu\\
&\vdots& \qquad \qquad \qquad\qquad \qquad \qquad \vdots
\nonu
\er
By equating the coefficients of $b_{2n+1}$ on both sides of
\rf{nicemkdvrel1} and \rf{nicemkdvrel2} one determines the parameters
$\xi_{2n+1}^{(\pm )}$. The first of them are  
\br
\xi_1^{(-)}&=&-\nu\qquad \qquad \qquad \;\,
\partial_x\xi_1^{(+)}=2\sinh^2\alpha
\lab{xirelmkdv}\\
\partial_x \xi_1^{(-)}&=&\frac{1}{2}\, q^2\qquad \qquad \qquad 
\partial_x\xi_3^{(+)}= \zeta_2^{(+)}\,\sinh\(2\, \alpha\)
\nonu\\
&\vdots&\qquad \qquad \qquad \qquad \qquad \quad \vdots
\nonu
\er

In order to construct the conserved charges one needs the time
component of the potentials to satisfy the boundary conditions
\rf{bc}. If one looks for solutions satisfying the conditions
\be
q\rightarrow  0\; ; \qquad \quad \partial_x^n q \rightarrow 0\; ;
\qquad \quad {\rm  as} \qquad x\rightarrow \pm \infty
\lab{bcmkdv}
\ee
then the potential \rf{mkdvpot} do satisfy \rf{bc},
i.e. $A_t\(t,x=\infty\)=A_t\(t,x=-\infty\)= -b_3$. From
\rf{zetarelmkdv} one observes that the parameters $\zeta_n^{(-)}$
depend locally on $q$ and its derivatives. Therefore  
\be
\zeta_n^{(-)}\rightarrow 0 \qquad \mbox{\rm and so} \qquad 
g_{-,F}  \rightarrow 1\qquad \quad {\rm as}\qquad x\rightarrow \pm
\infty
\ee
Therefore, according to the discussion below \rf{wgauge} one concludes
that the charges obtained from the potentials $A_{\mu}$, given in
\rf{mkdvpot},  and those from $a_{\mu}^{(-)}$, defined in
\rf{nicemkdvrel2},  are 
the same since they are related by a gauge transformation involving a
group element that goes to unity at spatial infinity.  

Assuming the conditions \rf{bcmkdv} one needs in addition that 
\be
\alpha\(t,x=\infty\)=\alpha\(t,x=-\infty\)
\lab{alphabcmkdv}
\ee
in order for the
potential ${\tilde A}_t$ to satisfy the boundary condition
\rf{bc}. However, notice that the mKdV equation \rf{mkdveq1} together
with the condition \rf{bcmkdv}  constitute a
conservation law which leads to the following conserved charge
\be
H_1=\int_{-\infty}^{\infty}dx\, q = 
-\left[\alpha\(t,x=\infty\)-\alpha\(t,x=-\infty\)\right]=
-\ln\frac{\tau_0}{\tau_1}\mid_{x=-\infty}^{x=\infty}
\ee
where we have used \rf{g0mkdvdef} and \rf{hirotataumkdv}. Therefore,
the conditions for the 
potential ${\tilde A}_{\mu}$ to give conserved charges imply that
$H_1$ should vanish. In addition, the condition \rf{alphabcmkdv} 
 is not sufficient for
the parameters $\zeta_n^{(+)}$ to vanish at spatial infinity, as seen from
\rf{zetarelmkdv}. Consequently, it does not guarantees that the
charges coming from ${\tilde A}_{\mu}$ and $a_{\mu}^{(+)}$ are the same, as
$g_{+,F}$ may not go to unity at spatial infinity. On the other hand,
the conditions for the potential $a_{\mu}^{(+)}$ to satisfy \rf{bc}
and so to lead to conserved charges, independently of what happens to
${\tilde A}_{\mu}$, is that
$\partial_t\xi_{2n+1}^{(+)}\(t,x=\infty\)= 
\partial_t \xi_{2n+1}^{(+)}\(t,x=-\infty\)$. However, that will
involve intricate conditions on $\alpha$. Therefore, the question if
one can construct conserved charges from the potentials  ${\tilde
  A}_{\mu}$ and $a_{\mu}^{(+)}$ depends on a very detailed analysis of
the boundary conditions satisfied by the solutions obtained by the
dressing method.  

The conditions \rf{bcmkdv} however, are suffucient to obtain conserved
charges from the potentials $A_{\mu}$ and $a_{\mu}^{(-)}$, as argued
above. Those charges are obtained following the discussion given in
\rf{wtgen}-\rf{wtvacgen}, and are given by 
\be
\Omega_{2n+1}^{(-)} = -\(2n+1\)\left[\xi_{2n+1}^{(-)}\(t,x=\infty\)-
  \xi_{2n+1}^{(-)}\(t,x=-\infty\)\right]\qquad\; n=0,1,2,\ldots 
\lab{mkdvnicecharges}
\ee
The asymptotic values of $\xi_{2n+1}^{(-)}$ can be evaluated using the
highest weight states of the fundamental representations of the
$sl(2)$ Kac-Moody algebra in a manner 
similar to that done for the sine-Gordon case in
\rf{xiplussg}-\rf{chargesminus}. 
 
Using \rf{xirelmkdv} one gets that the lowest charge is related to one
of the Hamiltonians of the mKdV hierarchy. Indeed, one has from
\rf{xirelmkdv}, \rf{mkdvnicecharges} and \rf{fieldtaurelmkdv} that 
\be
\Omega_{1}^{(-)} =-\frac{1}{2}\,\int_{-\infty}^{\infty}dx\, q^2 = \left(
  \nu\mid_{x=\infty}-\nu\mid_{x=-\infty}\right) = 
\frac{1}{2}\,\partial_x\ln\(\tau_0\,\tau_1\)\mid_{x=-\infty}^{x=\infty}
\ee
Therefore, we have here a situation very similar to the sine-Gordon
case (see \rf{energy-momentum-formula}) where the energy of the
solution is determined by the 
asymptotic behavior of the extra field associated to the central term
of the algebra. Our method therefore gives a very simple formula for the
energy, and also for the higher charges, of the mKdV solutions on the
orbit of the vacumm, $q^{\rm 
  vac}=\nu^{\rm vac}=0$, under the dressing transformation group.


\appendix

\section{The $sl(2)$ Kac-Moody algebra}
\label{sec:sl2km}
\setcounter{equation}{0}

The commutation relations of the $sl(2)$ Kac-Moody algebra are given
by \cite{goreview} 
\br
\sbr{T_3^m}{T_3^n}&=& \frac{1}{2}\, m \, \delta_{m+n,0}\, C, \nonu\\
\sbr{T_3^m}{T_{\pm}^n}&=&  \pm \, T_{\pm}^{m+n},\nonu\\
\sbr{T_+^m}{T_-^n}&=& 2\, T_3^{m+n} + m \, \delta_{m+n,0}\, C,
\er
where $C$ is the central term. 
The relevant basis for our calculations is given by
\be
b_{2m+1}= T_{+}^m + T_{-}^{m+1} \; ; \qquad F_{2m+1}= T_{+}^m -
T_{-}^{m+1} \; ; \qquad  
F_{2m} = 2\, T_3^{m} - \frac{1}{2} \, \delta_{m,0}\, C,
\ee
which satisfy 
\br
\sbr{b_{2m+1}}{b_{2n+1}} &=& \(2m+1\)\, \delta_{m+n+1,0}\, C,\nonu\\
\sbr{b_{2m+1}}{F_n} &=& -2\, F_{n+2m+1},\nonu\\
\sbr{F_{2m+1}}{F_{2n}} &=& - 2 \, b_{2(m+n)+1}, \nonu\\
\sbr{F_{2m+1}}{F_{2n+1}} &=& -(2m+1)\, \delta_{m+n+1,0}\, C,  \nonu\\
\sbr{F_{2m}}{F_{2n}} &=& 2m\, \delta_{m+n,0}\, C . 
\lab{newbasiscomrel}
\er
The indices of the generators correspond to the grades under
\be
\sbr{Q}{b_{2m+1}} = \(2m+1\)\,b_{2m+1} \; ; \qquad \qquad \quad
\sbr{Q}{F_n} = n\, F_n,
\ee
where
\be
Q= T_3^0+ 2 d \qquad \quad {\rm with} \qquad \quad \sbr{d}{T_i^m}= m
\, T_i^m \qquad i=3,+,- .
\lab{graddef}
\ee
In the case when the central term vanishes, {\it i.e.} $C=0$, the algebra
is called the $sl(2)$ loop algebra, and it admits finite matrix
representations. In the case of $2\times 2$ matrices on has
\br
b_{2m+1} = \(
\begin{array}{cc}
0& \lambda^m\\
\lambda^{m+1} & 0
\end{array}\), \quad 
F_{2m+1} = \(
\begin{array}{cc}
0& \lambda^m\\
-\lambda^{m+1} & 0
\end{array}\) , \quad 
F_{2m} = \lambda^m \(
\begin{array}{cc}
1& 0\\
0 & -1
\end{array}\) 
\lab{matrixrep}
\er
with $m=0,\pm 1,\pm 2\ldots$. In that case, the operator $d$ takes 
the form $d\equiv \lambda\,\frac{d\;}{d\,\lambda}$.

For $C\neq 0$ all the representations of the $sl(2)$ Kac-Moody
  algebra are infinite dimensional. The methods of constructing these 
  representations involve field theory techniques \cite{goreview},
  like the vertex operator given below. However, having $C\neq 0$
  leads to a very desirable property, namely the existence of the
  highest weight 
  state representations, {\it i.e.} representations that contain
  states that are 
  annihilated by positive root step operators (the generalization of
  $T_{+}$ in the algebra of angular momentum). Indeed, among the
  highest weight state representations of the $sl(2)$ Kac-Moody
  algebra there are two that play a very important role. They are the two
  fundamental representations, with highest weight states
  $\ket{\lambda_i}$, $i=0,1$, satisfying 
\br
T_3^0\, \ket{\lambda_0} &=& 0, \qquad \qquad \qquad T_3^0\, \ket{\lambda_1} =
\frac{1}{2}\,  \ket{\lambda_1},\nonumber \\
C\, \ket{\lambda_0} &=& \ket{\lambda_0}, \qquad \qquad\;\; 
C\, \ket{\lambda_1} = \, \ket{\lambda_1} 
\lab{highestweight1}
\er
and
\br
T_3^n\, \ket{\lambda_i} = T_{\pm}^n\, \ket{\lambda_i} =T_{+}^0\,
  \ket{\lambda_i} = 0,  \qquad \qquad n>0\; ; \quad i=0,1.
\lab{highestweight3}
\er
From \rf{highestweight1} one gets  
\be
F_0\, \ket{\lambda_0} = -\frac{1}{2}\, \ket{\lambda_0}, \qquad \qquad \qquad 
F_0\, \ket{\lambda_1} = \frac{1}{2}  \ket{\lambda_1}.
\lab{highestweight2}
\ee

An important mathematical tool in the study of solitons is the use of the
so-called vertex operator representations of the Kac-Moody algebras. In the
case of the sine-Gordon model the relevant representation is the one
involving the principal vertex operators. It is based on the Fock
space of the oscillators $b_{2n+1}$ satisfying the first relation in
\rf{newbasiscomrel} with $C=1$, {\it i.e.} 
\be
\sbr{b_{2m+1}}{b_{2n+1}} = \(2m+1\)\, \delta_{m+n+1,0}.
\ee
The vertex operator is defined as
\cite{goreview}: 
\be
V\(z\)\equiv :e^{{\cal Q}\(z\)}:= e^{{\cal Q}_{<}\(z\)}\,e^{{\cal
    Q}_{>}\(z\)},
\lab{vertexdef}
\ee
where, as usual, $:\;:$, denotes the normal ordering of the
oscillators ($b_{2n+1}$ with 
$n\geq 0$ are the annihilation operators, and the negative ones the
creation operators), and where
\be
{\cal Q}\(z\)\equiv {\cal Q}_{<}\(z\)+{\cal Q}_{>}\(z\) 
\ee
and
\be
{\cal Q}_{>}\(z\)= \sum_{n=0}^{\infty}\frac{2\, z^{-2n-1}}{2n+1}\,
b_{2n+1},\qquad \qquad  
{\cal Q}_{<}\(z\)= -\sum_{n=0}^{\infty}\frac{2\,z^{2n+1}}{2n+1}\, b_{-2n-1}
\ee
with $z$ being an arbitrary (complex) parameter. 

One can then show that in such a representation the generators $F_n$ are
given by \cite{goreview}: 
\be
F_n= \oint \frac{dz}{2\pi i z}\, z^n \, V\(z\).
\ee
There are two important   properties of the vertex operators which are
relevant for the solitons. First, the vertex operators are eigenstates
of the oscillators
\be
\sbr{b_{2n+1}}{V\(z\)}= -2\, z^{2n+1}\, V\(z\), \qquad \qquad n=0,\pm
1,\pm 2,\ldots
\lab{vertexeigen}
\ee
The second property is its operator product expansion 
\be
V\(z_1\)\,V\(z_2\) =\, :V\(z_1\)\,V\(z_2\): \,
\(\frac{z_1-z_2}{z_1+z_2}\)^2
\lab{vertexproduct2}
\ee
 so that $V\(z\)$ is nilpotent
\be
V\(z\)^2 = 0.
\lab{vertexnipotent}
\ee
One can also show that
$$
V\(z_1\)\,V\(z_2\)\,V\(z_3\)=\, :V\(z_1\)\,V\(z_2\)\,V\(z_3\):  
\, \(\frac{z_1-z_2}{z_1+z_2}\)^2\,\(\frac{z_1-z_3}{z_1+z_3}\)^2\, 
\(\frac{z_2-z_3}{z_2+z_3}\)^2
$$
and in general that
\be
\prod_{i=1}^n V\(z_i\)=\, :\prod_{i=1}^n V\(z_i\): \prod_{i<j=1}^{n}
\(\frac{z_i-z_j}{z_i+z_j}\)^2.
\lab{vertexproductn}
\ee
We also have that
\be
\bra{\lambda_0}:\prod_{i=1}^n V\(z_i\): \ket{\lambda_0}=1,\qquad
\mbox{and} \qquad  
\bra{\lambda_1}:\prod_{i=1}^n V\(z_i\): \ket{\lambda_1}=(-1)^n.
\lab{vertexvev}
\ee

\vspace{2cm}

\noindent {\bf Acknowledgements:} LAF is partially supported by a CNPq
grant. WJZ acknowledge a FAPESP grant supporting his visit to
IFSC/USP.

\end{document}